\documentclass[aps,prl,reprint,twocolumn,superscriptaddress,english,nolongbibliography]{revtex4-2}
\usepackage{amssymb}
\usepackage{graphicx}
\usepackage{amsmath}
\usepackage{epsfig}
\usepackage[latin9]{inputenc}
\usepackage{array}
\usepackage{multirow}
\usepackage{color}
\usepackage{esint}
\usepackage{bm}
\usepackage{bbm}
\usepackage[bookmarks=false,pdftex,colorlinks=true,urlcolor=black,linkcolor=black,citecolor=black,breaklinks=true]{hyperref}
\usepackage{babel}
\usepackage{epstopdf}
\usepackage{mathtools}
\usepackage{soul}
\usepackage[dvipsnames]{xcolor}
\usepackage{tikz}

\usepackage{setspace} 
\newcommand{\push}{\hspace{0.05cm}}
\newcommand{\pull}{\hspace{-0.05cm}}

\begin{document}
	
	\setcounter{equation}{0} \setcounter{figure}{0}
	\setcounter{table}{0} \setcounter{page}{1} \makeatletter
	\title{Realizing discontinuous quantum phase transitions in a strongly-correlated driven optical lattice}
	
	\author{Bo Song}
	\affiliation{Cavendish Laboratory, University of Cambridge, J.J. Thomson Avenue, Cambridge CB3 0HE, United Kingdom\looseness=-1}
	\author{Shovan Dutta}
	\affiliation{Cavendish Laboratory, University of Cambridge, J.J. Thomson Avenue, Cambridge CB3 0HE, United Kingdom\looseness=-1}
	\author{Shaurya Bhave}
	\affiliation{Cavendish Laboratory, University of Cambridge, J.J. Thomson Avenue, Cambridge CB3 0HE, United Kingdom\looseness=-1}
	\author{Jr-Chiun Yu}
	\affiliation{Cavendish Laboratory, University of Cambridge, J.J. Thomson Avenue, Cambridge CB3 0HE, United Kingdom\looseness=-1}
	\author{Edward Carter}
	\affiliation{Cavendish Laboratory, University of Cambridge, J.J. Thomson Avenue, Cambridge CB3 0HE, United Kingdom\looseness=-1}
	\author{Nigel Cooper}
	\affiliation{Cavendish Laboratory, University of Cambridge, J.J. Thomson Avenue, Cambridge CB3 0HE, United Kingdom\looseness=-1}
    \affiliation{Department of Physics and Astronomy, University of Florence, Via G. Sansone 1, 50019 Sesto Fiorentino, Italy\looseness=-1}
	\author{Ulrich Schneider}
	\affiliation{Cavendish Laboratory, University of Cambridge, J.J. Thomson Avenue, Cambridge CB3 0HE, United Kingdom\looseness=-1}
	
	\date{\today}
	
	\begin{abstract}
		Discontinuous quantum phase transitions and the associated metastability play central roles in diverse areas of physics ranging from ferromagnetism to the false vacuum decay in the early universe. Using strongly-interacting ultracold atoms in an optical lattice,  we realize a driven many-body system whose quantum phase transition can be tuned from continuous to discontinuous.
		Resonant shaking of a one-dimensional optical lattice hybridizes the lowest two Bloch bands, driving a novel transition from a Mott insulator to a $\pi$-superfluid, i.e., a superfluid state with staggered  phase order.
		For weak shaking amplitudes, this transition is discontinuous (first-order) and the system can remain frozen in a metastable state, whereas for strong shaking, it undergoes a continuous transition toward a $\pi$-superfluid.
		Our observations of this metastability and hysteresis are in good quantitative agreement with numerical simulations and pave the way for exploring the crucial role of quantum fluctuations in discontinuous transitions.
	\end{abstract}
	
	\maketitle
	\newpage

	Phase transitions are ubiquitous in physics, ranging from thermal phenomena such as the boiling of water to magnetic transitions in solids, and from cosmological phase transitions in the early universe~\cite{kibble1980some} to the transition into a quark-gluon plasma in high-energy collisions~\cite{mclerran1986physics}. Particularly intriguing are quantum phase transitions that occur at temperatures close to absolute zero and are driven by quantum rather than thermal fluctuations~\cite{sachdev2011}. So far, the focus has been on continuous quantum phase transitions (second- or higher-order), such as most magnetic or superfluid-to-Mott-insulator transitions. There is however a renewed interest in discontinuous (first-order) quantum phase transitions, which are characterized by an inherent metastability: The system can remain in its initial phase after crossing the transition~\cite{lifshitz1972quantum}. 
	Of particular interest is to understand the {\it quantum} decay of such a metastable state, termed false vacuum decay, which is relevant in particle physics and cosmology as an analogue of the `Big Bang' in inflationary universes~\cite{coleman1977fate,vilenkin1983birth,fialko2015fate,fialko2017universe,ng2021fate}. 
	
	Ultracold atoms in optical lattices provide a pristine and controllable platform to investigate quantum phases and phase transitions in isolated many-body systems. 
	While continuous phase transitions have been studied extensively in these systems~\cite{greiner2002quantum, gross2017quantum}, discontinuous transitions have so far been limited to weakly-interacting condensates~\cite{struck2013engineering,trenkwalder2016quantum,campbell2016magnetic,qiu2020observation}, for which the quantum decay of a metastable state is strongly suppressed: It requires the collective tunneling of all atoms within a healing length, leading to an exponential reduction~\cite{fialko2017universe}, analogous to the suppression of tunneling in large-spin systems~\cite{OWERRE20151}. 
	
    Here, we engineer a discontinuous quantum phase transition from a strongly-interacting Mott insulator to a superfluid in a resonantly driven one-dimensional (1D) optical lattice~\cite{eckardt2017colloquium}. This is analogous to the spin-1/2 quantum XY model in which quantum fluctuations play a significant role~\cite{osterloh2002scaling}. 
    Our approach is based on periodically modulating or shaking the position of the lattice. In contrast to earlier work that relied on the modification of tunneling matrix elements by strong off-resonant shaking~\cite{eckardt2005superfluid,zenesini2009coherent,eckardt2017colloquium,michon2018phase}, we employ resonant drives that couple the lowest two Bloch bands.
    In non- or weakly-interacting lattice systems, off-resonant shaking is central to generating topological band structures~\cite{cooper2019topological}, and resonant shaking can result in spontaneous symmetry breaking \cite{zheng2014strong} analogous to a ferromagnetic quantum phase transition \cite{parker2013direct}.
     
    In the present work, the parameters of the undriven lattice are chosen such that the many-body ground state in the lowest band is a Mott insulator (MI), but the far stronger tunneling in the first excited band alone would result in a superfluid state. We term this state a $\pi$-superfluid ($\pi$-SF), as the negative sign of the tunneling implies that condensation occurs at the edge of the Brillouin zone, i.e., with a staggered phase order. 
    In the presence of a near-resonant drive, the corresponding dressed bands will cross, and we can drive the MI-to-$\pi$-SF transition by ramping the shaking frequency and thereby shifting the relative energies of the bands.
    Crucially, this transition is discontinuous at weak coupling strengths, because the non-staggered correlations in the MI are incompatible with the staggered order of the $\pi$-SF~\cite{lim2008staggered, strater2015orbital}. For stronger coupling strengths, on the other hand, the bands are strongly hybridized and the transition becomes continuous.

	\begin{figure}
		\includegraphics[width=1\linewidth]{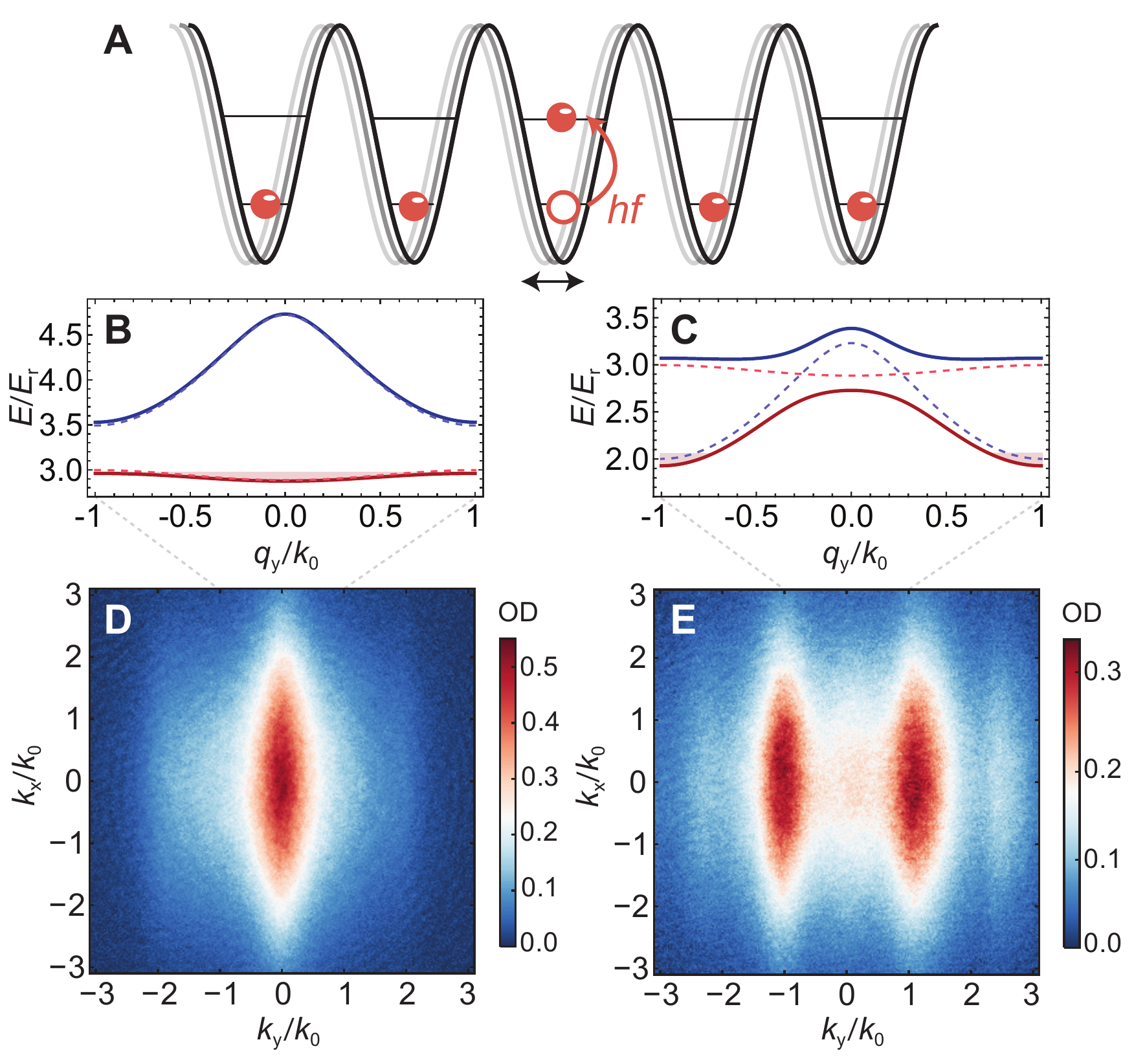}
		\centering
		\caption{\textbf{Schematic of the shaken lattice}. (\textbf{A})
		Resonant modulation (shaking) of the lattice position hybridizes the lowest two single-particle Bloch bands. (\textbf{B}) and (\textbf{C}) show the lowest two bands in the Mott insulating (off-resonant) and $\pi$-superfluid (resonant) regimes with shaking parameters $(f,\mathcal{A})=(15\text{ kHz},11.5\text{ nm})$ and $(f,\mathcal{A})=(21.5\,\text{kHz},11.5\,\text{nm})$, respectively. Dashed lines indicate the bare (uncoupled) bands.
		While the momentum distribution of the Mott insulator in the lowest band is centered around zero, the $\pi$-superfluid is quasi-condensed at the Brillouin zone boundary.
		(\textbf{D}) and (\textbf{E}) show average time-of-flight images ($t_{tof}=16\,$ms, OD: Optical density) of atoms released from the shaken lattice corresponding to (\textrm{B}) and (\textrm{C}), respectively, following the direct sweep protocol depicted in Fig.~\ref{fig:phase}A.}
		\label{fig:schematic}
	\end{figure}
	
	\paragraph{Shaken lattice.}
	In our experiment, a Bose-Einstein condensate of around $10^5$ rubidium ($^{87}$Rb) atoms is adiabatically loaded into 
	a three-dimensional simple-cubic lattice with lattice depths of $(V_x, V_y, V_z) = (33, 8.4, 49)\,E_r$, where $E_r=\frac{\hbar^2k_0^2}{2m}$ denotes the recoil energy,  $m$ the mass of $^{87}$Rb, $k_0=\frac{2\pi}{\lambda}$, and $\lambda=726\,$nm is the wavelength of the lattice light. The lattices along $x$ and $z$ are sufficiently deep such that tunneling along those directions is negligible and the cloud is partitioned into independent tubes along the $y$ direction, forming independent 1D Mott insulators. The atoms are held in this static lattice for $10$-$15\,$ms before we start modulating (shaking) the position of the $y$-lattice by sinusoidally changing the corresponding laser frequency with modulation frequency $f$ using an acousto-optical modulator in double-pass configuration. Due to the retro-reflected setup, this translates to a spatial shaking of the lattice at the atoms' position. The shaking amplitude $\mathcal{A}$ and resulting coupling strength $\Omega \propto f^2\, \mathcal{A}$ between the lowest two bands are determined by the frequency modulation depth and the distance ($l_0\approx45\,$cm) between the atoms and the retro reflector~\cite{SuppMat}. As the size of the atomic cloud is negligible compared to $l_0$, the shaking amplitude remains constant across the cloud. 
	
	In contrast to off-resonant shaking schemes \cite{zenesini2009coherent, michon2018phase, weinberg2015multiphoton}, typical amplitudes in this resonant case are tiny, namely less than $4\%$ of the lattice constant $d=\lambda/2$, such that
	the dressed and bare bands (solid and dashed lines in Fig.~\ref{fig:schematic}B) essentially coincide away from resonance.  In the resonant case, on the other hand, the two bands hybridize and thereby strongly increase the bandwidth of the relevant dressed band, shown in red in Fig.~\ref{fig:schematic}C.
	
	\paragraph{Extended Bose-Hubbard model.} 
	In the lattice frame, the shaking gives rise to an oscillatory force in the $y$ direction \cite{Arimondo2012}. Expanding the field operator in terms of Wannier functions in the lowest two bands and moving to the rotating frame, one finds an effective Hamiltonian for sufficiently deep lattices~\cite{SuppMat},
	\begin{flalign}
	    \nonumber \hat{H} =& \sum_{\langle i,j \rangle} J_b \hat{b}_i^{\dagger} \hat{b}_j 
	                        -J_a \hat{a}_i^{\dagger} \hat{a}_j 
                            + \sum_{j} \Delta \hat{b}_j^{\dagger} \hat{b}_j 
                            + \frac{\Omega}{2} \big( \hat{a}_j^{\dagger} \hat{b}_j + \hat{b}_j^{\dagger} \hat{a}_j  \big) &&\\
        & + \sum_{j} \frac{U_{a}}{2} \hat{a}_j^{\dagger} \hat{a}_j^{\dagger} \hat{a}_j \hat{a}_j 
                   + \frac{U_{b}}{2} \hat{b}_j^{\dagger} \hat{b}_j^{\dagger} \hat{b}_j \hat{b}_j
                   + V \hat{a}_j^{\dagger} \hat{b}_j^{\dagger} \hat{b}_j \hat{a}_j \;,\hspace{-0.2cm} &&
    \label{simple_model}
	\end{flalign}
	where $\hat{a}_j$ and $\hat{b}_j$ represent bosonic annihilation operators in the ground and first excited uncoupled bands, respectively, $J_b \gg J_a$ are the nearest-neighbor tunneling in the two bands, and $U_{a,b}$ and $V$ are the intraband and interband on-site interactions. The effective detuning $\Delta = \Delta_E - h f$ is measured with respect to the average gap $\Delta_E$ between the lowest two bands and $h$ is Planck's constant. In the numerical simulations based on the density-matrix renormalization group (DMRG)~\cite{Schollwoeck2011}, we also add next-nearest-neighbor tunneling in the excited band (which is comparable to $J_a$) and corrections of order $1/f$ from Floquet theory \cite{goldman2014periodically,SuppMat}. 
    Additionally, we have performed three-band simulations which show that higher bands are off-resonant and unimportant for the used experimental sweeps~\cite{SuppMat}.  While higher bands would become significant for much slower large-amplitude sweeps, they can be suppressed using superlattice techniques~\cite{strater2015orbital}.
 
	\paragraph{Phases and their signatures.} 
	The scenario in Eq.~\eqref{simple_model} was first discussed in Ref.~\cite{strater2015orbital} and can be interpreted as a frustrated ladder model, as the tunneling elements in the two bands have opposite signs. Consequently, a many-body state can either satisfy the links in the ground band or the links in the excited band, but not both at the same time. This frustrated hopping is at the heart of the discontinuous phase transition. For weak coupling strengths in the discontinuous regime, either the ground or the excited band is occupied almost exclusively depending on the detuning $\Delta$. In the ground band, the interaction energy dominates over the kinetic energy and the system forms a Mott insulator with small and positive nearest-neighbor correlations $\langle \hat{a}_i^{\dagger} \hat{a}_{i+1}\rangle>0$. In the excited band, in contrast, the larger bandwidth implies that kinetic energy dominates and the system is in a $\pi$-superfluid state, where the negative sign of the hopping results in a staggered order with $\langle \hat{b}_i^{\dagger} \hat{b}_{i+1}\rangle<0$. As there can be no continuous transition connecting these two incompatible orders, the system has to choose one of them, giving rise to a discontinuous transition. Conversely, strong coupling results in strongly hybridized bands, where the transition from positive to negative nearest-neighbor correlations (from non-staggered to staggered) happens already within the Mott insulator regime. 
	The phase transition now smoothly connects a staggered Mott insulator to a staggered $\pi$-superfluid and is of the normal continuous Kosterlitz-Thouless type.
	
	\begin{figure}
		\includegraphics[width=1\linewidth]{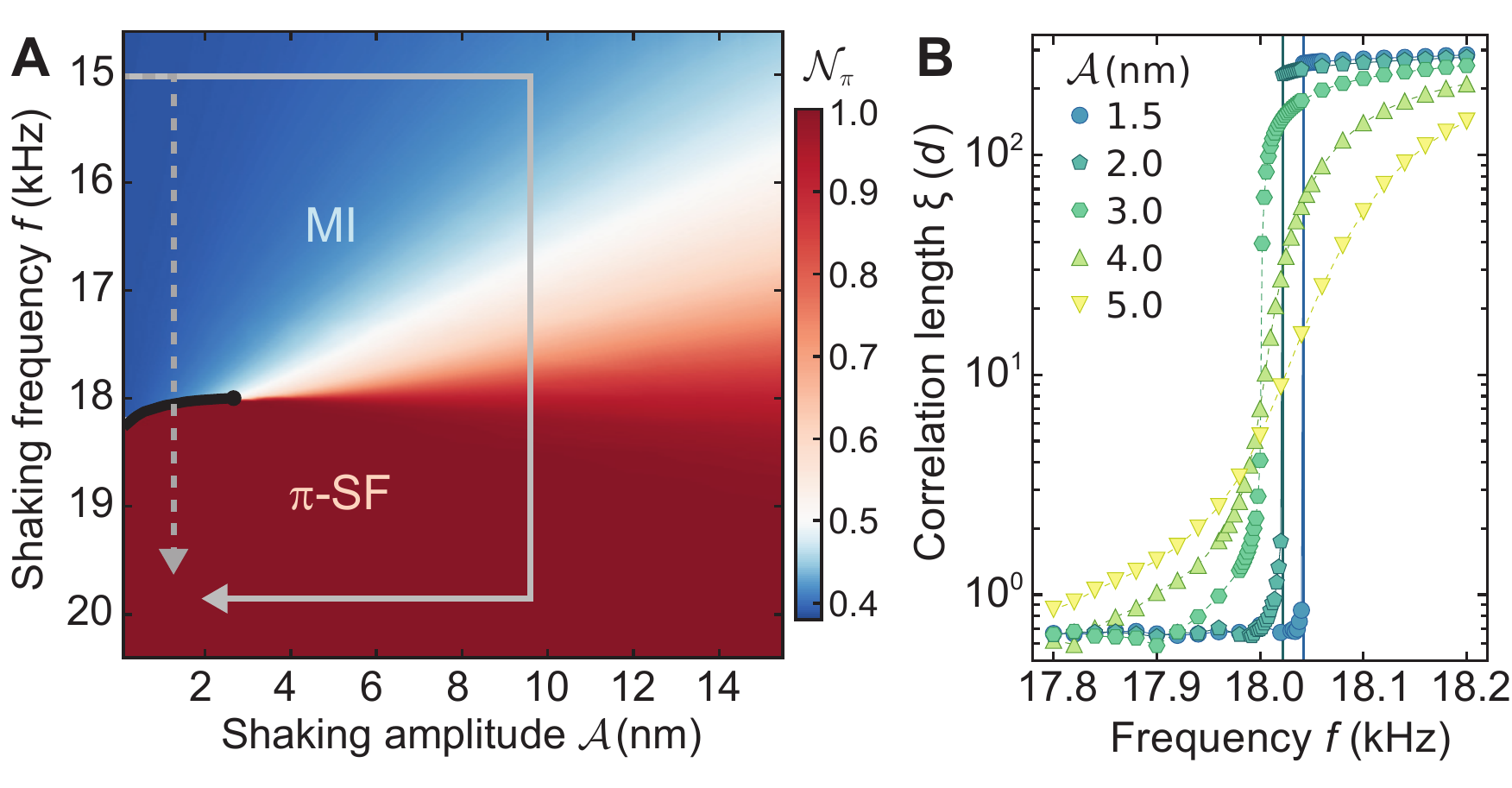}
		\centering
		\caption{\textbf{Simulated phase diagram of the shaken lattice}. (\textbf{A}) Normalized occupation $\mathcal{N}_\pi$ around momentum $\pm \hbar k_0$. For small shaking amplitudes $\mathcal{A}$, Mott insulator (blue) and $\pi$-superfluid (red) phases are separated by an abrupt discontinuous transition (black solid line), which turns continuous for $\mathcal{A} \gtrsim 2.8\,$nm. Gray arrows indicate the direct (dashed) and indirect (solid) sweeps used in Fig.~\ref{fig:phasediag}.  (\textbf{B}) The distinction between the two types of phase transitions is also reflected by the ground-state correlation length $\xi$. In contrast to weak amplitudes, where $\xi$ jumps discontinuously at the phase boundary, it varies smoothly  for large $\mathcal{A}$. Here, $\xi$ is calculated by averaging $\langle \hat{b}_i^{\dagger} \hat{b}_j \rangle$, obtained from a two-band DMRG with 64 sites and unity filling~\cite{SuppMat}, over the bulk.
	    }
		\label{fig:phase}
	\end{figure}
	
	These phases exhibit distinct signatures in the momentum distribution measured by time-of-flight (TOF) imaging, as shown in Figs.~\ref{fig:schematic}D,E.
	In the initially prepared (non-staggered) Mott insulator, the broad momentum distribution is centered around $k=0$ (Fig.~\ref{fig:schematic}D) and the absence of satellite peaks demonstrate the absence of long-range coherence. Conversely, in a $\pi$-SF, the atoms are concentrated around $k=\pm k_0$ (Fig.~\ref{fig:schematic}E), and the relatively narrow satellite peaks at higher momenta signal the presence of at least short-range coherence. To distinguish between these phases, we extract the normalized population at the band edge, $\mathcal{N_\pi} := n_\pi/(n_0+n_\pi)$, by counting the number of atoms in fixed windows around $k=0$ (for $n_0$) and $k=\pm k_0$ (for $n_{\pi}$)~\cite{SuppMat}.

	
	\paragraph*{Phase diagram.}
	Figure~\ref{fig:phase}A shows the simulated phase diagram (for the effective Hamiltonian) at unity filling.  For weak shaking amplitudes, $\mathcal{N_\pi}$ changes abruptly with shaking frequency $f$, indicating the discontinuous phase transition. Beyond a critical drive amplitude of $\mathcal{A}_c\simeq2.8\,$nm, the change in $\mathcal{N_\pi}$ becomes smooth as the phase transition turns continuous. 
	This change from discontinuous to continuous is further corroborated by the vanishing of the jump in the simulated correlation lengths (Fig.~\ref{fig:phase}B) and the behavior of the entanglement entropy presented in the Supplementary Material~\cite{SuppMat}.
	
	\begin{figure}
		\includegraphics[width=1\linewidth]{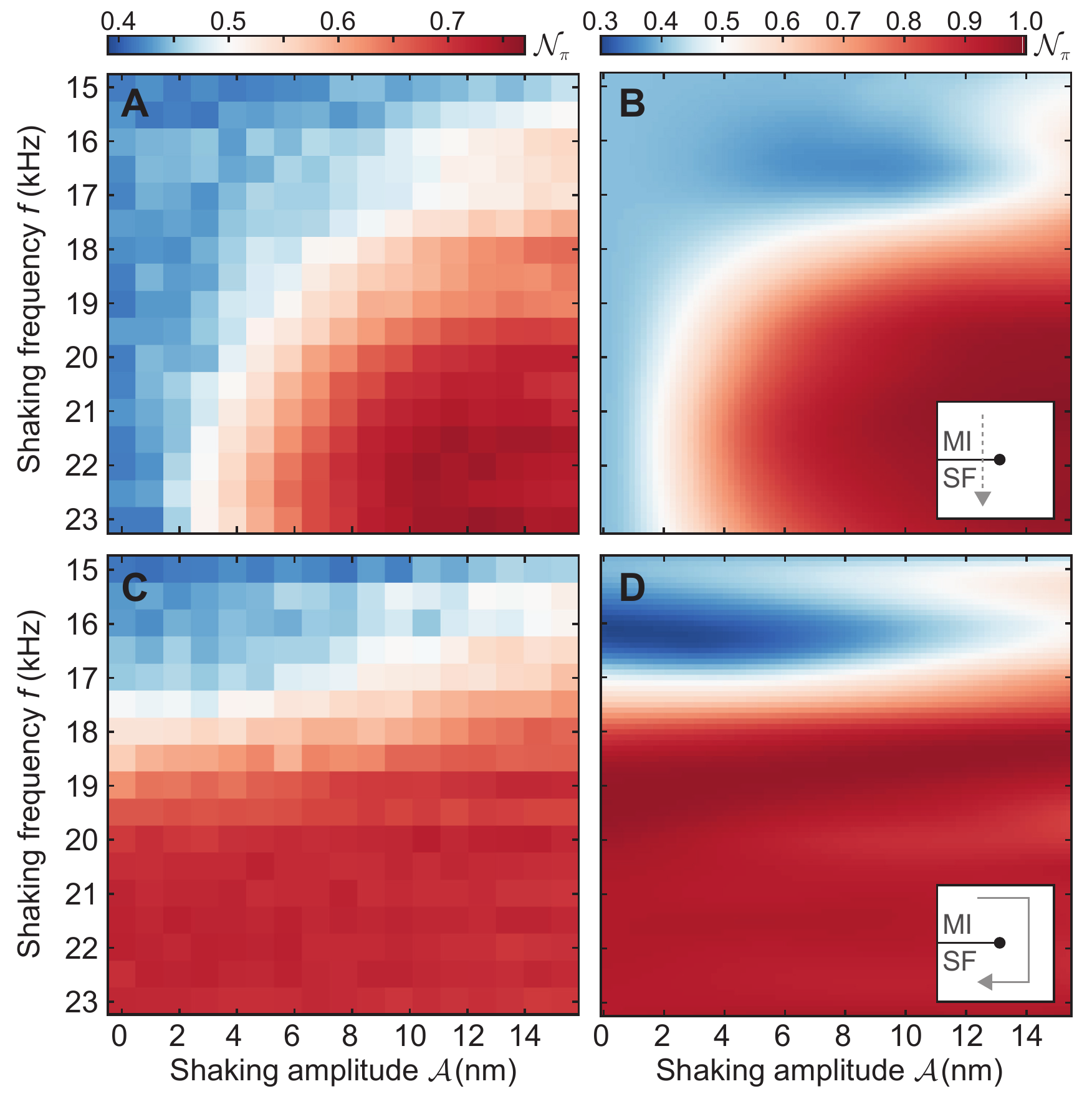}
		\centering
		\caption{\textbf{Direct and indirect sweeps across phase transitions}. (\textbf{A}) and (\textbf{B}) show, respectively, the experimental and simulated band-edge populations $\mathcal{N}_\pi$ following the direct sweeps (inset and dashed arrow in Fig.~\ref{fig:phase}A)  as a function of the final amplitude and frequency. (\textbf{C}) and (\textbf{D}) show the corresponding results for indirect sweeps (solid arrow). Direct and indirect sweeps broadly agree in the continuous regime at large amplitudes, but differ crucially in the discontinuous regime: Whereas indirect sweeps always pass through the continuous phase transition where the occupation broadly follows the ground state, the observed populations hardly change during direct sweeps through the discontinuous transition, highlighting the associated metastability. 
		Simulations are based on adaptive tDMRG \cite{Spaeckel2019} with the full time-dependent two-band Hamiltonian for 10 sites and unity filling~\cite{SuppMat}.}
		\label{fig:phasediag}
	\end{figure}
	
	\paragraph{Dynamics of phase transition and quantum metastability.}
	The discontinuous nature and the associated metastability are revealed by comparing two different frequency sweeps termed direct and indirect sweeps that are indicated by arrows in Fig.~\ref{fig:phase}A. In the direct sweep, the shaking amplitude is linearly increased from 0 to the final amplitude in $125\,\mu$s at a fixed off-resonant frequency of $15\,$kHz before the shaking frequency is increased linearly to the final frequency in $600\,\mu$s. 
	In the indirect sweep, first the shaking amplitude is linearly increased from 0 to a common large amplitude of $\mathcal{A}=9.6\,$nm in $125\,\mu$s at $15\,$kHz. Then the shaking frequency is linearly swept to the final frequency in $600\,\mu$s before the shaking amplitude is decreased to the final amplitude in $300\,\mu$s. The indirect sweep protocol ensures that the system always undergoes a continuous phase transition, circumventing the discontinuous regime. For these sweeps, the atoms largely remain in the original of the two dressed bands (red solid line in Figs.~\ref{fig:schematic}B,C) and, for sufficiently slow sweeps, stay close to the ground state~\cite{clark2016universal}.

	Figure~\ref{fig:phasediag} shows the state at the end of the sweeps as a function of the final amplitude and frequency. While the results for the indirect sweeps (Figs.~\ref{fig:phasediag}C,D) agree well with the  static phase diagram of Fig.~\ref{fig:phase}A, the direct sweeps (Figs.~\ref{fig:phasediag}A,B) clearly demonstrate the metastable nature of the discontinuous phase transition: In the discontinuous regime, the system remains in the initial Mott insulating state characterized by small $\mathcal{N_\pi}$, even though the ground state changes to a $\pi$-SF. 
	
\begin{figure}
		\includegraphics[width=1\linewidth]{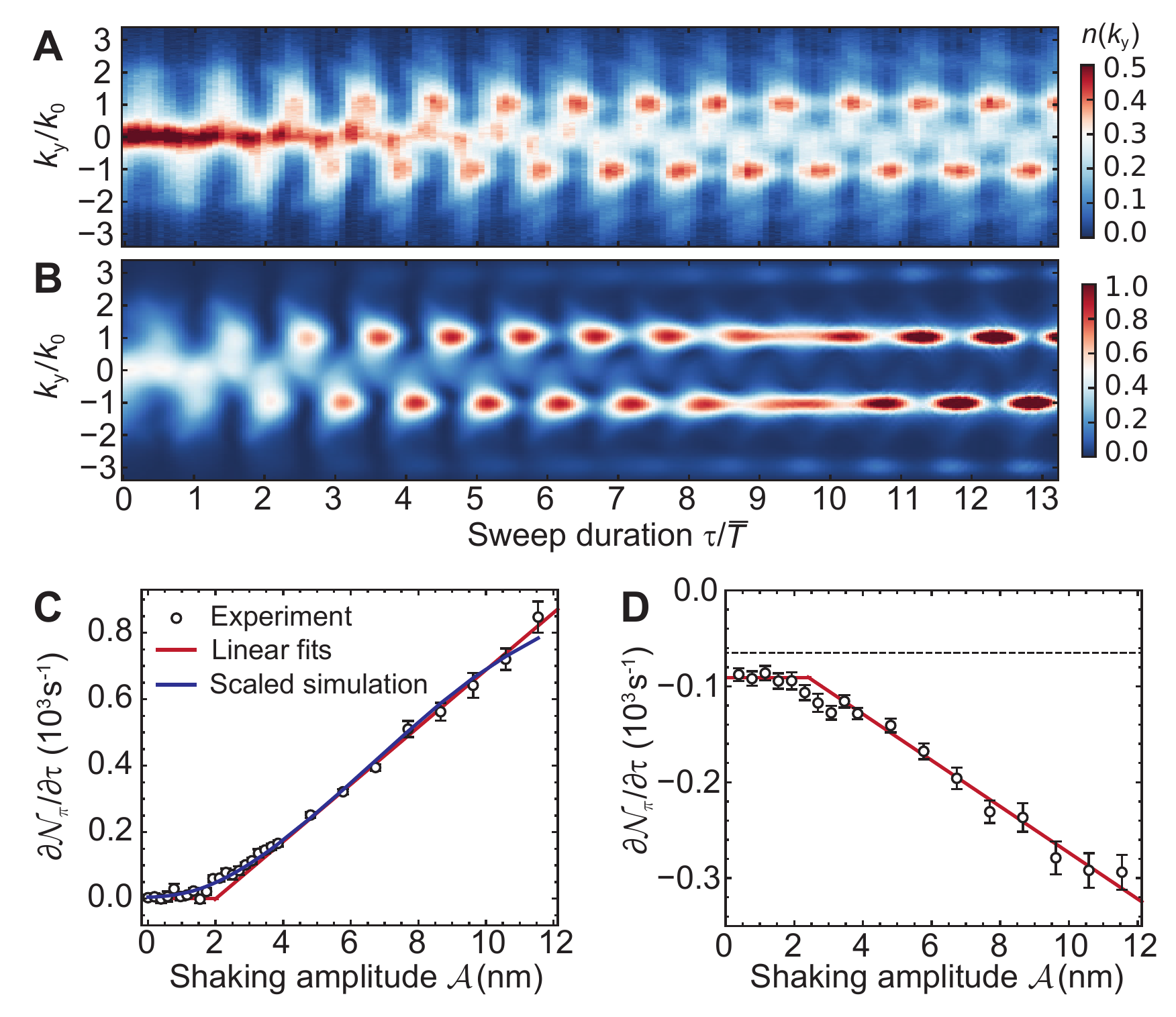}
		\centering
		\caption{\textbf{Dynamics of phase transitions}. (\textbf{A}) and (\textbf{B}) show normalized experimental and simulated momentum distributions $n(k_y)$, respectively, after a direct frequency sweep from the Mott insulating ($15\,$kHz) to the $\pi$-superfluid regime ($21\,$kHz) with $\mathcal{A}=11.5\,$nm, as a function of the sweep duration $\tau$. One can see a buildup of pronounced peaks around $\pm k_0$ along with the induced micromotion. (\textbf{C}) The initial growth rate of the peaks, $\partial\mathcal{N}_\pi/\partial \tau$, at different shaking amplitudes. In contrast to strong shaking where the rate increases linearly with $\mathcal{A}$, for weak amplitudes the populations are nearly frozen after crossing the discontinuous phase transition, exhibiting metastability. The red solid line represents a piecewise linear fit and the blue solid line shows the simulation scaled empirically by $0.45$ to account for the experimental inhomogeneity. (\textbf{D}) $\partial\mathcal{N}_\pi/\partial \tau$ for backward sweeps, starting from a $\pi$-superfluid ($21\,$kHz) prepared via the continuous transition back to the Mott regime ($15\,$kHz). It also shows metastability for small $\mathcal{A}$, resulting in hysteresis. Black dashed line indicates an incoherent background decay of the $\pi$-superfluid~\cite{SuppMat}.
		}
		\label{fig:phasetrans}
	\end{figure}

	Next, we investigate the dynamics of the phase transition by varying the duration of the direct sweeps. Figure~\ref{fig:phasetrans}A shows the normalized final momentum distribution $n(k_y)$ after a sweep through the continuous transition from a MI state ($f_i=15\,$kHz) into the $\pi$-SF regime  ($f_f=21\,$kHz) as a function of the sweep duration $\tau$, expressed in units of the average drive period $\bar{T}=2\pi/\bar{f}$, where $\bar{f}:= (f_i+f_f)/2$. With increasing $\tau$, prominent peaks emerge at $k_y = \pm k_0$ corresponding to the Brillouin zone boundary. The oscillation between $+k_0$ and $-k_0$ stems from the micromotion in the accelerated lattice~\cite{goldman2014periodically,eckardt2017colloquium} combined with Bragg reflections at the band edge, and the oscillation frequency is equal to $\bar{f}$~\cite{SuppMat}.
	The observed dynamics, which in the experiment is averaged over multiple tubes, is in good qualitative agreement with the tDMRG simulation at unity filling in Fig.~\ref{fig:phasetrans}B.
	
	We repeat this sweep measurement for different shaking amplitudes (see~\cite{SuppMat} for data) and extract the initial growth rate of the band-edge population, $\partial\mathcal{N}_\pi/\partial \tau$, see Fig.~\ref{fig:phasetrans}C.  The vanishing rate of change at weak amplitudes highlights the metastability associated with the discontinuous transition - the system remains in the initial state,  even though the phase transition has been crossed. The observed rates are in good agreement with the simulation up to an overall scaling factor that we attribute mainly to the inhomogeneity of the dipole trap. 
	We partially attribute the finite susceptibility around the edge of the discontinuous regime to the presence of nucleation points due to initial entropy and boundary effects.

	In Fig.~\ref{fig:phasetrans}D, we additionally explore backward sweeps where the system is initially prepared in the $\pi$-SF regime ($f_i=21\,$kHz) via the continuous phase transition (solid arrow in Fig.~\ref{fig:phase}A) and then swept back to the MI regime ($f_f=15\,$kHz). While the overall structure  is very similar to the forward (MI-to-$\pi$-SF) sweeps in Fig.~\ref{fig:phasetrans}C, we find an additional small nonzero decay rate that is slightly larger than the independently measured incoherent dephasing rate $ \Gamma\sim -65\,s^{-1}$ (dashed line in Fig.~\ref{fig:phasetrans}D~\cite{SuppMat}), which we attribute to excitations in the prepared $\pi$-SF state that could act as nucleation centers. 
	Both forward and backward sweeps demonstrate the metastability and hysteresis associated with the discontinuous transition.\newline

	\noindent\textbf{Conclusion} Using a resonant shaking scheme, we realize a tunable quantum phase transition in an isolated strongly-correlated lattice system and demonstrate the strikingly different dynamics of discontinuous and continuous transitions. Ramping across the discontinuous transition reveals the associated quantum metastability, in good agreement with tensor-network simulations.  
	Resonantly driven lattice systems open a new avenue for engineering novel quantum phases and studying genuinely quantum discontinuous transitions and other intriguing critical phenomena. Our technique can be directly extended to higher dimensions where numerical simulations are unfeasible. Future studies can also investigate the decay mechanism of metastable many-body states and explore the emergence of spatial structures resulting from quantum fluctuations and the influence of nucleation points. This will shed light on the dynamics and structure formation in the early universe~\cite{ng2021fate}. Resonantly driven lattices will furthermore enable novel studies on the dynamical scaling across such transitions~\cite{qiu2020observation, shimizu2018dynamics, pelissetto2018out, coulamy2017dynamics} and open the door to studying prevalent first-order transitions in interacting topological phases~\cite{amaricci2015first,zhao2018crystallization,wu2021z2} using quantum simulators.
	\newline
	
	\noindent \textbf{Acknowledgments} This work was partly funded by the European Commission ERC Starting Grant \mbox{QUASICRYSTAL}, the EPSRC Grant EP/R044627/1
	and Programme Grant \mbox{DesOEQ} (EP/P009565/1), and by a Simons Investigator Award.
	We are grateful to Emmanuel Gottlob and Andr\'e Eckardt for fruitful discussions.
	
\begingroup
\renewcommand{\addcontentsline}[3]{}
\renewcommand{\section}[2]{}

\endgroup

\onecolumngrid
\clearpage
\setstretch{1.2}
\begin{center}
\textbf{\large Supplementary Materials: Realizing discontinuous quantum phase transitions \\ in a strongly-correlated driven optical lattice}
\end{center}

\setcounter{equation}{0}
\setcounter{figure}{0}
\setcounter{table}{0}
\setcounter{secnumdepth}{3}
\makeatletter

\renewcommand{\thefigure}{S\arabic{figure}}
\renewcommand{\theequation}{S\arabic{equation}}
\renewcommand{\bibnumfmt}[1]{[S#1]}
\renewcommand{\citenumfont}[1]{S#1}
\renewcommand{\theHfigure}{S\thefigure}
\bfseries
\tableofcontents
\normalfont

\section{Modulation of lattice frequency}\label{sec:modfreq}

A BEC of $1\times 10^5$ $^{87}$Rb atoms in the $|F=1$, $m_F=-1\rangle$ state is adiabatically loaded into a 3D lattice with depths of $(V_x, V_y, V_z) = (33, 8.4, 49)E_r$. All lattice depths are increased exponentially from zero to their final values in $45\,$ms with a time constant of $10\,$ms. This results in a Mott insulating state. After a hold time of $10$-$15\,$ms,  we begin to modulate (shake) the position of the $y$-lattice by sinusoidally changing the corresponding laser frequency. To this end,  we modulate the frequency of the radio frequency wave feeding an acousto-optic modulator (AOM) in double-pass configuration. The resulting frequency of the lattice laser $f_l(t)$ becomes
\begin{equation}\label{eqLattMod}
f_l (t) = f_c + A_l \sin(2\pi f t),
\end{equation}
where $f_c$ is the central frequency, $A_l$ the frequency modulation depth and $f$ denotes the modulation frequency. For a given fixed distance $l_0$ between the atoms and the retro mirror, this frequency modulation results in the desired modulation of the lattice position~\cite{folling2007direct}, namely $y(t)=y_0+\mathcal{A}\sin(2\pi f t)$, where $y_0$ is the original lattice position and the modulation depth $\mathcal{A}$ denotes the maximum displacement of the lattice in space. For the fixed distance between atoms and the retro-reflector $l_0=45\,$cm and lattice spacing $d=\lambda/2=363\,$nm in our setup, a frequency modulation amplitude of $A_l=1\,$MHz corresponds to a modulation amplitude of $\mathcal{A} = 0.33\times10^{-2}\,d = 1.2\,$nm in space. Since the size of the BEC is very small compared to $l_0$, the inhomogeneity of the shaking amplitude $\mathcal{A}$ across the cloud is less than 0.01\%. 

\begin{figure}[h]
	\includegraphics[width=0.5\linewidth]{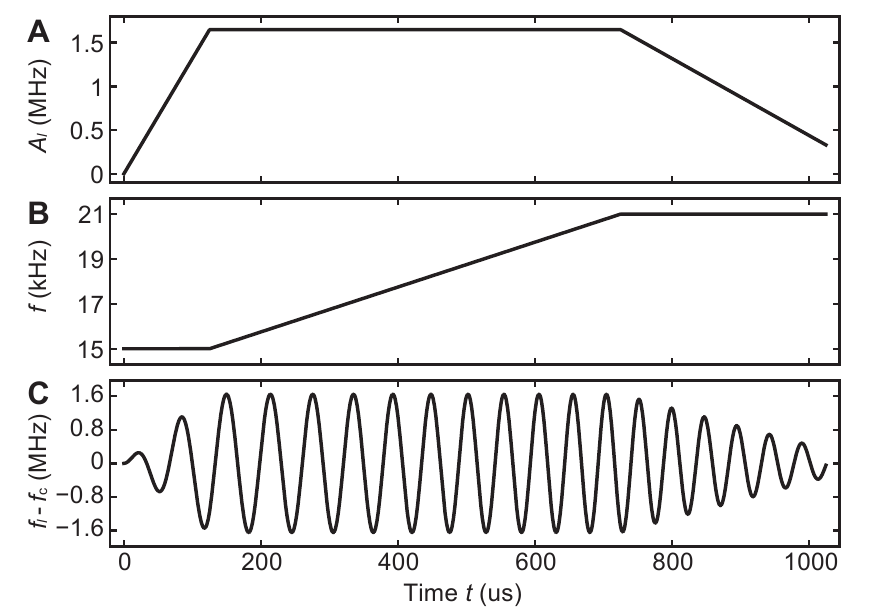}
	\centering
	\caption{\textbf{Frequency modulation of lattice laser}. An example of the indirect sweep sequence with a final shaking frequency $f_f=21\,$kHz and a final shaking amplitude $A_f=0.33\,$MHz. (\textbf{A}), (\textbf{B}) and (\textbf{C}) show the variation of the shaking amplitude, shaking frequency and laser frequency, respectively.}
	\label{figs1}
\end{figure}

The use of a double-pass AOM enables us to control both the shaking amplitude as well as the shaking frequency in a straightforward manner using a radio frequency source that is based on a  direct digital synthesis (DDS) chip. Here we discuss one example of the indirect sweep protocol used in Fig.~\ref{fig:phasediag} in the main text. We ramp the amplitude from zero to an intermediate value $A_m=1.65\,$MHz in $t_1=125\,\mu$s and then the amplitude is kept constant for the next $600\,\mu$s until $t_2=725\,\mu$s. Finally the amplitude is linearly changed to the final value, e.g., $A_f=0.33\,$MHz at $t_3=1025\,\mu$s, shown in Fig.~\ref{figs1}A.
\begin{equation}
A_l(t)= 
\begin{cases}
A_m\frac{t}{t_1},		&  t_1 > t \geq 0 \\
A_m,     	&  t_2 > t \geq t_1 \\
(A_f-A_m)\frac{t-t_2}{t_3-t_2} + A_m		&  t_3 \geq t \geq t_2
\end{cases}.
\end{equation}

\noindent The corresponding frequency variation is given by
\begin{equation}
	f(t)= 
	\begin{cases}
	f_i,		&  t_1 > t \geq 0 \\
	(f_f-f_i)\frac{t-t_1}{t_2-t_1} + f_i     &  t_2 > t \geq t_1 \\
	f_f,		&  t_3 \geq t \geq t_2
	\end{cases},
\end{equation}
where the initial shaking frequency during the first $125\,\mu$s is $f_i=15\,$kHz. Then the frequency is swept linearly in time (linear chirp) to a final value, e.g., $f_f=21\,$kHz in $600\,\mu$s, see Fig.~\ref{figs1}B. In order to guarantee a phase continuous waveform, we apply phase modulation instead of direct frequency control. In this case, Eq.~\eqref{eqLattMod} becomes $f_L(t) = f_c + A_l\sin(\phi(t))$ and the required phase profile is given by  
\begin{equation}
    \phi(t)=\int_{0}^{t}2\pi f(t) dt,
\end{equation}
where $f(t)$ denotes the time-dependent modulation frequency and we set the initial phase $\phi_0=0$. The final time-dependent laser frequency in the indirect sweep is shown in Fig.~\ref{figs1}C. 

\clearpage
\section{Micromotion}\label{sec:micromotion}
In the lattice frame, the acceleration of the lattice gives rise to a periodic oscillation of all quasimomenta. For atoms within a given band and with quasimomenta away from the Brillouin zone boundary, this micromotion is not observable in the lab frame, as it is precisely compensated by the transformation between the two frames of reference. For quasimomenta close to $\pm\hbar  k_0$, however, the situation changes: Even a small-amplitude oscillation around the Brillouin zone boundary will give rise to Bragg reflections that transfer atoms between $+\hbar k_0$ and $-\hbar k_0$, which are mapped to opposite real momenta during time-of-flight imaging, thereby resulting in a strong oscillation of the center-of-mass position $(\propto \bar{k}_y)$ observed after time of flight.

\begin{figure}[h]
	\includegraphics[width=1\linewidth]{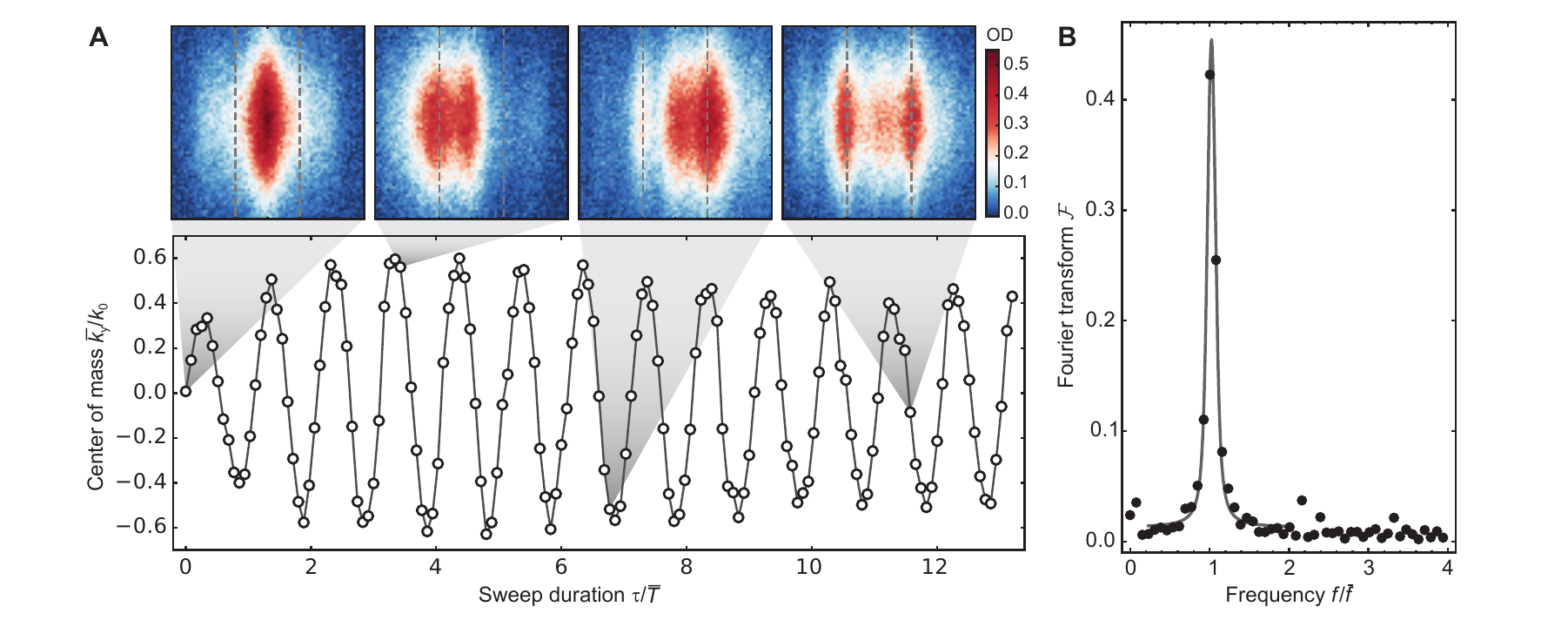}
	\centering
	\caption{\textbf{Micromotion}. (\textbf{A}) Due to the combination of micromotion and Bragg reflections, the center of mass in momentum space oscillates as a function of the sweep duration $\tau$. $\bar{T}=1/\bar{f}$ denotes the average drive period with the average shaking frequency $\bar{f}$. The insets show examples of time-of-flight pictures for different sweep durations, highlighting the Bragg reflections between $\pm\hbar k_0$: In the lab frame, the population oscillates between these two discrete momenta, with equal populations at $\pm \hbar k_0$ coinciding with the times when the lattice has zero displacement. Dashed lines indicate $\pm \hbar k_0$. (\textbf{B}) The Fourier transform shows that the oscillation frequency is equal to the average shaking frequency $\bar{f}$. The line is a fit by a Voigt profile.}
	\label{figs3}
\end{figure}

For a fixed modulation frequency $f$, the resulting oscillation in $\bar{k}_y$ would have the same frequency, i.e.\ $\bar{k}_y\propto \cos(2\pi f t)$. 
For more general sweeps, the final phase of the modulation can be written as $\phi(\tau)=2\pi\int_{0}^{\tau} f(t) dt$, where the initial phase offset is set to $\phi(0)=0$. In the sweep measurements, we vary the sweep duration $\tau$ during which the shaking frequency is linearly ramped from the initial frequency $f_i$ to the final frequency $f_f$. This results in $\phi(\tau)=2\pi\, \bar{f}\tau$, i.e. the final phase oscillates with the average shaking frequency $\bar{f}=(f_i+f_f)/2$. We experimentally confirm this by extracting the center of mass in momentum space from the normalized momentum distributions $n(k_y)$ shown in Fig.~\ref{fig:phasetrans}A in the main text according to $\bar{k}_y = \int k_y~n(k_y)dk_y$. In Fig.~\ref{figs3}, $\bar{k}_y$ is plotted as a function of sweep duration $\tau$, and  the oscillation frequency determined by a Voigt fit to its Fourier transform, $f_{peak} = 1.02\bar{f}$, is found to be equal to $\bar{f}$ within the experimental uncertainty.

\clearpage
\section{Band-edge population $\mathcal{N}_\pi$}

\begin{figure}[h]
	\includegraphics[width=1\linewidth]{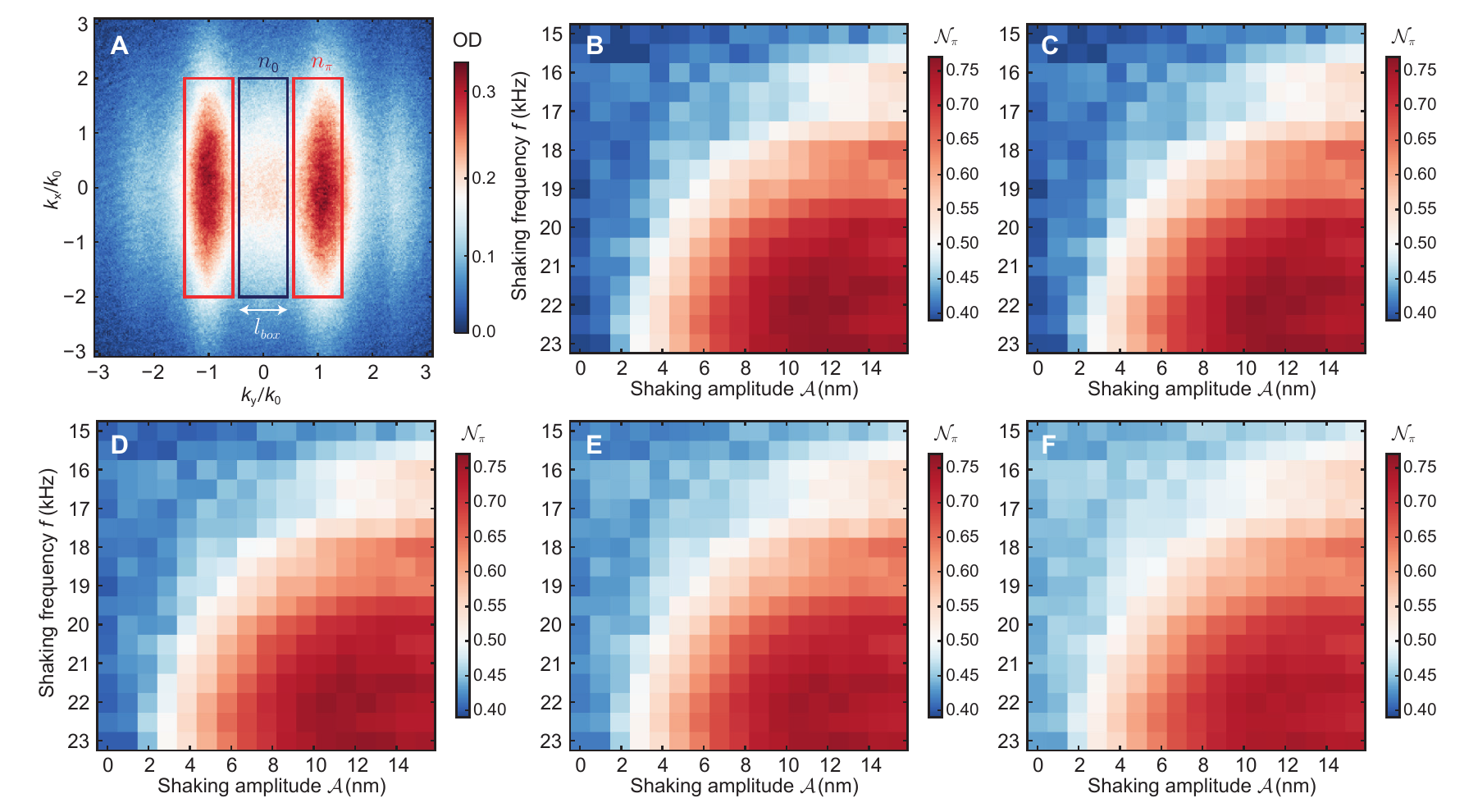}
	\centering
	\caption{\textbf{The band-edge population $\mathcal{N}_\pi$.} (\textbf{A}) Atom numbers around $k=0$ ($n_0$) and $k=\pm k_0$ ($n_\pi$) are counted inside the boxes in blue and red, respectively. Note that $n_{\pi}$ is the sum over the two red boxes. (\textbf{B}-\textbf{F}) are measured with different box sizes $l_{box}/k_0=0.1,0.2,0.4,0.6,$ and $0.8$, respectively.}
	\label{figs2}
\end{figure}

We measure the (normalized) at the band edge population $\mathcal{N}_\pi := n_\pi/(n_0+n_\pi)$ by directly counting the  numbers of atoms in fixed boxes around  $k_y=0$ (for $n_0$) and  $k_y=\pm \pi$ (for $n_\pi$), as shown in Fig.~\ref{figs2}A. The image is averaged over 30 individual images, demonstrating the stability of peak positions. The same data is shown in Fig.~\ref{fig:schematic}E in the main text. To further evaluate the robustness of this method, we vary the width of these boxes ($l_{box}$) along the modulated $k_y$  direction. Figures~\ref{figs2}(B-F) show the extracted populations following the direct sweeps from Fig.~\ref{fig:phasediag} in the main paper for box widths $l_{box}$ ranging from $0.1 k_0$ to $0.8k_0$. While the absolute value of $\mathcal{N}_\pi$ is modestly dependent on the used box size, the observed pattern is robust and insensitive to the used box width. In Figs.~\ref{fig:phasediag} and \ref{fig:phasetrans} in the main text, $l_{box} = 0.6k_0$.

\section{Forward and backward sweeps}

\begin{figure}[h]
		\includegraphics[width=1.0\linewidth]{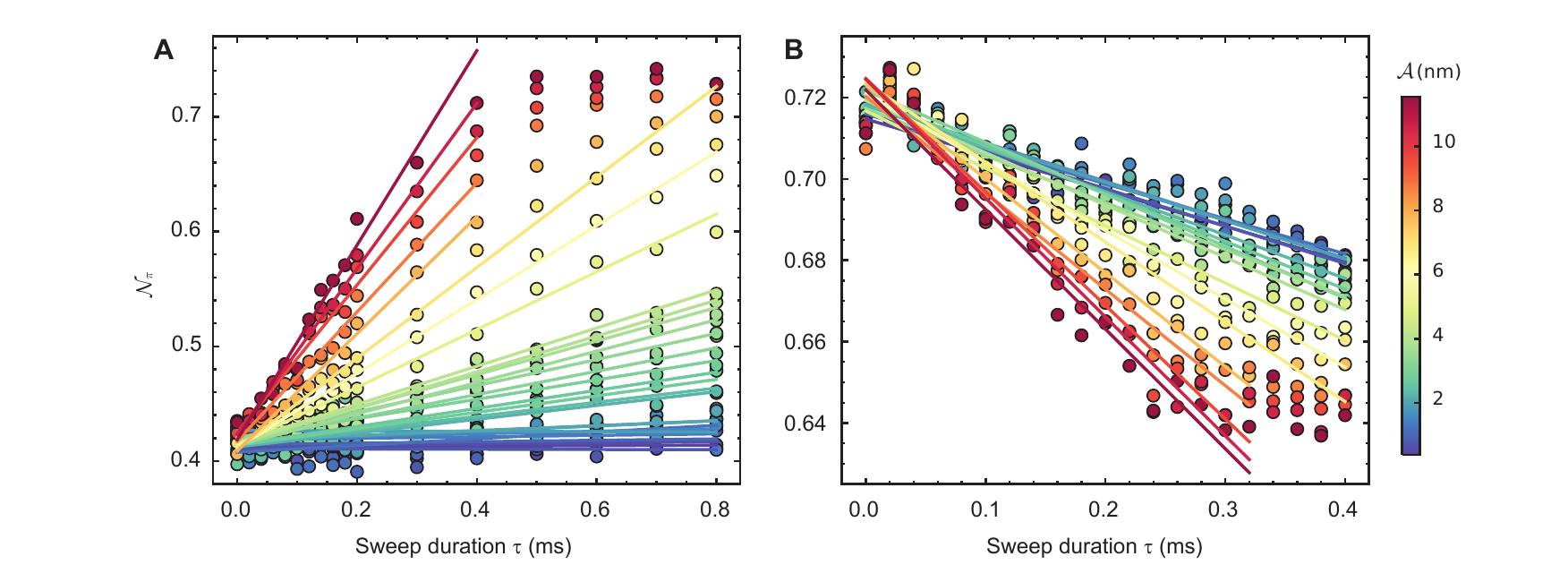}
		\centering
		\caption{\textbf{The band-edge population after different sweep durations for different shaking amplitudes}. (\textbf{A}) and (\textbf{B}) show the final population $\mathcal{N}_\pi$ after direct sweeps in the forward (from a MI to a $\pi$-SF state) and backward (from a $\pi$-SF to a MI state) directions. Solid lines are linear fits.}
		\label{figs:phasetrans}
	\end{figure}

We measure the final population $\mathcal{N}_\pi$ after direct forward sweeps (from a MI to a $\pi$-SF) and direct backward sweeps (from a $\pi$-SF to a MI). In the forward sweeps, the system is initially prepared in a MI regime ($f_i=15\,$kHz) and we measure  $\mathcal{N}_\pi$ as a function of the duration $\tau$ of the frequency sweep to the final shaking frequency ($f_f=21\,$kHz), see Fig.~\ref{figs:phasetrans}A. In the backward sweeps, the system is first prepared in a $\pi$-SF regime ($f_i=21\,$kHz) following the indirect sweep protocol described in Sec.~\ref{sec:modfreq}, before sweeping the shaking frequency back to $f_f=15\,$kHz with varying duration $\tau$, see Fig.~\ref{figs:phasetrans}B. 
In both cases, we  extract $\partial\mathcal{N}_\pi/\partial \tau$ by the linear fits shown as solid lines. The fit results are shown in Figs.~\ref{fig:phasetrans}(C-D) in the main text.

\section{Dephasing of superfluid correlations}

\begin{figure}[h]
	\includegraphics[width=0.5\linewidth]{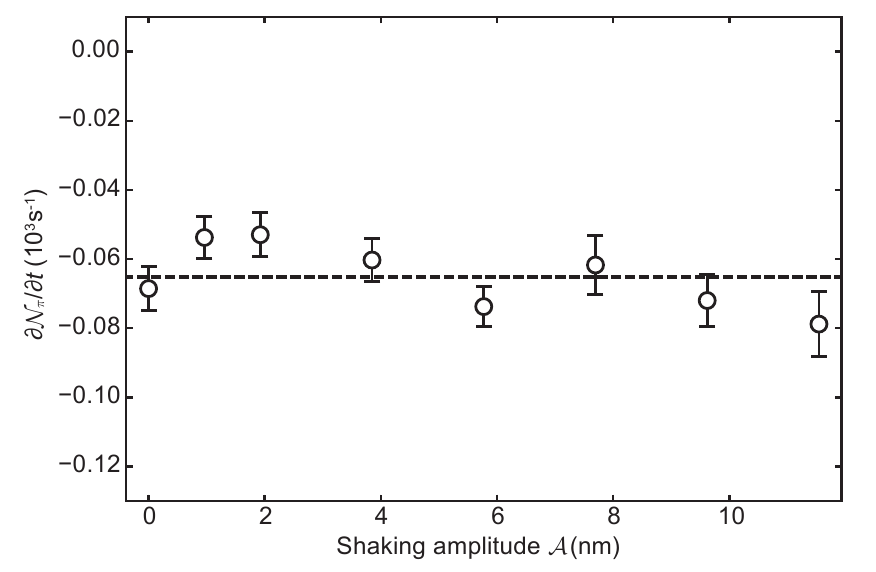}
	\centering
	\caption{\textbf{Dephasing of superfluid correlations}. Rate of change of the band-edge population with respect to hold time, $\partial\mathcal{N}_\pi/\partial t$, measured for different shaking amplitudes $
	\mathcal{A}$. Error-bars indicate the stand error of each linear fit. The dashed line indicates the average decay rate.}
	\label{figs4}
\end{figure}

To understand the finite value of $\partial\mathcal{N}_\pi/\partial \tau$ at weak shaking amplitudes in the backward sweep (from $\pi$-SF back to MI in Fig.~\ref{figs:phasetrans}B and Fig.~\ref{fig:phasetrans} in the main text), we investigate the lifetime of the $\pi$-superfluid correlations in the shaken lattice. We prepare the system in the $\pi$-SF regime at a shaking frequency of 21kHz with different shaking amplitudes via the indirect sweep and then measure the evolution of  $\mathcal{N}_\pi$ during a hold time $t$.  We apply a linear fit to extract $\partial\mathcal{N}_\pi/\partial t$ and, as shown in Fig.~\ref{figs4}, find $\partial\mathcal{N}_\pi/\partial t=-65\,s^{-1}$, irrespective of the shaking amplitude $\mathcal{A}$. This incoherent dephasing  produces a nonzero offset for $\partial\mathcal{N}_\pi/\partial \tau$ in the backward sweeps in Fig.~\ref{fig:phasetrans} in the main text and is indicated there by the dashed line.

\clearpage

\section{\label{sec:model}Extended Hubbard model}

The shaken lattice is described by an effective 1D Hamiltonian
\begin{equation}
\hat{H} (t) = 
\int \pull dy \push \hat{\psi}^{\dagger}(y) \left[-\frac{\hbar^2}{2m} \partial_y^2 + V_{\text{lat}} (y - s (t)) \right] \hat{\psi}(y) 
+ \frac{g_{\text{1D}}}{2} \int \pull dy \push \hat{\psi}^{\dagger}(y) \hat{\psi}^{\dagger}(y) \hat{\psi}(y) \hat{\psi}(y) \;,
\end{equation}
where $\hat{\psi}(y)$ is the boson field operator, $m$ is the mass, $V_{\text{lat}}(y) = V_0 \sin^2 (\pi y / d)$ is the unperturbed lattice potential with lattice depth $V_0$ and lattice spacing $d$, $s(t)$ is the displacement due to shaking, and $g_{\text{1D}}$ is an effective interaction strength. Transforming to the lattice frame as in Ref.~\cite{sArimondo2012}, one finds
\begin{equation}
\hat{H}(t) = 
\int \pull dy \push \hat{\psi}^{\dagger}(y) \left[-\frac{\hbar^2}{2m} \partial_y^2 + V_{\text{lat}} (y) + F(t) y\right] \hat{\psi}(y) 
+ \frac{g_{\text{1D}}}{2} \int \pull dy \push \hat{\psi}^{\dagger}(y) \hat{\psi}^{\dagger}(y) \hat{\psi}(y) \hat{\psi}(y) \;,
\label{suppeq:latframehamil}
\end{equation}
where $F(t) = m \ddot{s}(t)$ is an inertial force. For a periodic shaking with constant amplitude $\mathcal{A}$ and frequency $\omega \equiv 2\pi f$, $F(t) = m \omega^2 \mathcal{A} \cos \omega t$ (up to a phase). The shaking in the experiment resonantly couples the lowest two bare bands of the lattice. Thus, we approximate the field operator as
\begin{equation}
\hat{\psi} (y) = 
\sum_{j} w_1(y-y_j)\push \hat{a}_j + w_2 (y-y_j)\push \hat{b}_j \push,
\end{equation}
where $w_1$ and $w_2$ are the Wannier functions centered at sites $y_j$ in the lowest and first excited bands, respectively, and $\hat{a}_j$ and $\hat{b}_j$ are the corresponding bosonic annihilation operators. Substituting the above expansion into Eq.~\eqref{suppeq:latframehamil} and retaining the most significant terms for a deep lattice, we obtain (up to a constant energy shift)
\begin{align}
\nonumber \hat{H}(t) =
 &\; \Delta_E \sum_j \hat{b}_j^{\dagger} \hat{b}_j 
 - J_a \sum_{\langle i,j \rangle} \hat{a}_i^{\dagger} \hat{a}_j 
 + J_b \sum_{\langle i,j \rangle} \hat{b}_i^{\dagger} \hat{b}_j 
 + J_b^{\prime} \sum_{\langle\langle i,j \rangle \rangle} \hat{b}_i^{\dagger} \hat{b}_j 
 + F(t) d \Big [ \sum_j j (\hat{a}_j^{\dagger} \hat{a}_j + \hat{b}_j^{\dagger} \hat{b}_j ) 
 + \alpha_{ab} \sum_j \hat{a}_j^{\dagger} \hat{b}_j + \hat{b}_j^{\dagger} \hat{a}_j \Big] \\
& + \frac{U_a}{2} \sum_j \hat{a}_j^{\dagger} \hat{a}_j^{\dagger} \hat{a}_j \hat{a}_j 
+ \frac{U_b}{2} \sum_j \hat{b}_j^{\dagger} \hat{b}_j^{\dagger} \hat{b}_j \hat{b}_j 
+ U_{ab} \sum_j \hat{a}_j^{\dagger} \hat{b}_j^{\dagger} \hat{b}_j \hat{a}_j 
+ \frac{U_{ab}}{4} \sum_j (\hat{a}_j^{\dagger} \hat{a}_j^{\dagger} \hat{b}_j \hat{b}_j + \text{h.c.}) \;,
\label{suppeq:hubbard1}
\end{align}
where $\Delta_E:=\varepsilon_b - \varepsilon_a$, $\varepsilon_{a,b}$ are the average energies of the two bands, $\langle i,j \rangle$ denotes nearest neighbors, $\langle\langle i,j \rangle\rangle$ denotes next-nearest neighbors, and $\alpha_{ab}$ is the coupling amplitude between the two bands,
\begin{equation}
\alpha_{ab} = \frac{1}{d}\int \pull dy \push y \push w_1(y) w_2(y) \;.
\end{equation}
The tunneling and interaction energies can be calculated from the dispersion and Wannier functions of the energy bands, which can be solved exactly in terms of Mathieu functions for a sinusoidal lattice \cite{sDrese1997}. Note that the Wannier functions can be taken to be real valued. The experiment is performed with ${}^{87}$Rb atoms with $d=363$~nm and $V_0 = 8.4 E_r$ where $E_r= 4.36$ kHz is the recoil energy (we list all energies in units of frequency). The transverse motion is frozen due to strong transverse lattice depths of $33 E_r$ and $49 E_r$. These parameters give $\Delta_E = 19.7$~kHz, $J_a = 0.12$~kHz, $J_b = 1.29$~kHz, $J_b^{\prime} = 0.17$~kHz, $U_a = 2.88$~kHz, $U_b = 1.73$~kHz, $U_{ab} = 2.36$~kHz, and $\alpha_{ab} = 0.15$. 

For a single-band Hubbard model, the Mott-superfluid transition is known to occur at $J/U \approx 0.3$ \cite{sEjima2012}. Thus, in our setup, the ground state of the lower band is a Mott insulator ($J_a/U_a \approx 0.04$), whereas that of the upper band is a $\pi$-superfluid ($J_b/U_b \approx 0.74$). The experiment starts with all atoms in the lower band and excites the upper band by sweeping the shaking frequency through resonance. To simulate these sweeps, we evolve with the time-dependent Hamiltonian in Eq.~\eqref{suppeq:hubbard1} with 10 particles on 10 sites using an adaptive tDMRG routine \cite{spaeckel2019time}. We use a fourth-order Euler stepper with a time step of 0.05 $\mu$s, keeping singular values above $10^{-10}$, with a maximum bond dimension of 100 and up to 4 particles per site, for which the results converge. The above approach can also be extended to incorporate corrections from higher bands, e.g., see Sec.~\ref{sec:3band} for simulation of sweeps including the third (second excited) band.

For a constant periodic drive $F(t) = m \omega^2 \mathcal{A} \cos \omega t$, one can find a simpler description in the rotating frame via the transformation $\hat{R} = e^{{\rm i} \omega t \sum_j \hat{b}_j^{\dagger} \hat{b}_j}$, which gives $\hat{R} \hat{b}_j \hat{R}^{\dagger} = e^{-{\rm i} \omega t} \hat{b}_j$ and $\hat{R}\big({\rm i}\hbar\partial_t - \hat{H}(t)\big)\hat{R}^{\dagger} \equiv {\rm i}\hbar\partial_t - \hat{H}_{\text{rot}}$, where $\hat{H}_{\text{rot}}$ is the rotating-frame Hamiltonian. Using Eq.~\eqref{suppeq:hubbard1}, one obtains
\begin{equation}
 \hat{H}_{\text{rot}} =
 \hat{H}_0 + \big(\hat{H}_1 e^{{\rm i}  \omega t} + \hat{H}_2 e^{{\rm i}  2\omega t} + \text{h.c.} \big)\;,
 \label{suppeq:Hrot}
\end{equation}
with
\begin{align}
\hat{H}_0 =
&\; \Delta \sum_j \hat{b}_j^{\dagger} \hat{b}_j 
- J_a \sum_{\langle i,j \rangle} \hat{a}_i^{\dagger} \hat{a}_j 
+ J_b \sum_{\langle i,j \rangle} \hat{b}_i^{\dagger} \hat{b}_j 
+ J_b^{\prime} \sum_{\langle\langle i,j \rangle \rangle} \hat{b}_i^{\dagger} \hat{b}_j 
+ \frac{\Omega}{2} \sum_j \hat{a}_j^{\dagger} \hat{b}_j + \hat{b}_j^{\dagger} \hat{a}_j \\
& + \frac{U_a}{2} \sum_j \hat{a}_j^{\dagger} \hat{a}_j^{\dagger} \hat{a}_j \hat{a}_j 
+ \frac{U_b}{2} \sum_j \hat{b}_j^{\dagger} \hat{b}_j^{\dagger} \hat{b}_j \hat{b}_j 
+ U_{ab} \sum_j \hat{a}_j^{\dagger} \hat{b}_j^{\dagger} \hat{b}_j \hat{a}_j \;, \\
\hat{H}_1 =
&\; \frac{F_0 d}{2} \sum_j j (\hat{a}_j^{\dagger} \hat{a}_j + \hat{b}_j^{\dagger} \hat{b}_j) \;, \\
\hat{H}_2 =&\;
 \frac{\Omega}{2} \sum_j \hat{b}_j^{\dagger} \hat{a}_j 
 + \frac{U_{ab}}{4} \sum_j \hat{b}_j^{\dagger} \hat{b}_j^{\dagger} \hat{a}_j \hat{a}_j \;,
\end{align}
where $\Delta := \Delta_E - \hbar \omega$ is the detuning from the averaged band gap, $\Omega := F_0 d \alpha_{ab}$ is the Rabi frequency, and $F_0 := m \omega^2 \mathcal{A}$ is the maximum force. For a deep lattice, the terms within parentheses in Eq.~\eqref{suppeq:Hrot} represent fast oscillations, since $\omega$ is close to the band gap. One can find an effective (Floquet) Hamiltonian that generates the time-averaged dynamics by using a high-frequency Magnus expansion, as derived in Ref. \cite{sgoldman2014periodically}. For $\omega \to \infty$, this is simply $\hat{H}_0$. For finite $\omega$, the most significant correction comes from the linear potential gradient $\hat{H}_1$, which renormalizes the hopping amplitudes. Retaining all other terms of order $1/\omega$, we find
\begin{align}
\nonumber \hspace{-0.25cm} \hat{H}_{\text{eff}} = 
& - J_a \mathcal{J}_0(\eta) \sum_{\langle i,j \rangle} \hat{a}_i^{\dagger} \hat{a}_j 
+ J_b \mathcal{J}_0(\eta) \sum_{\langle i,j \rangle} \hat{b}_i^{\dagger} \hat{b}_j
+ J_b^{\prime} \mathcal{J}_0(2\eta) \sum_{\langle\langle i,j \rangle\rangle} \hat{b}_i^{\dagger} \hat{b}_j 
+ \sum_{j} \Delta_{\text{eff}} \hspace{0.05cm} \hat{b}_j^{\dagger} \hat{b}_j 
+ \frac{\Omega}{2} \big( \hat{a}_j^{\dagger} \hat{b}_j + \hat{b}_j^{\dagger} \hat{a}_j  \big) \\
& + \sum_{j} \frac{U_a^{\text{eff}}}{2} \hat{a}_j^{\dagger} \hat{a}_j^{\dagger} \hat{a}_j \hat{a}_j 
+ \frac{U_b^{\text{eff}}}{2} \hat{b}_j^{\dagger} \hat{b}_j^{\dagger} \hat{b}_j \hat{b}_j 
+ U_{ab} \hspace{0.05cm} \hat{a}_j^{\dagger} \hat{b}_j^{\dagger} \hat{b}_j \hat{a}_j
+ \frac{U_{ab}^2}{8\hbar\omega} \hspace{0.05cm} \hat{n}^a_j \hat{n}^b_j (\hat{n}^b_j - \hat{n}^a_j) 
+ \frac{\Omega U_{ab}}{8\hbar\omega} \left[ \hat{a}_j^{\dagger} (\hat{n}^b_j - \hat{n}^a_j) \hat{b}_j + \text{h.c.} \right] ,
\end{align}
where $\eta := F_0 d/ (\hbar \omega)$, $\Delta_{\text{eff}} := \Delta + \Omega^2/ (4\hbar \omega)$, $U_a^{\text{eff}} := U_a - U_{ab}^2/(8\hbar\omega)$, $U_b^{\text{eff}} := U_b + U_{ab}^2/(8\hbar\omega)$, $\hat{n}^a_j := \hat{a}_j^{\dagger} \hat{a}_j$, $\hat{n}^b_j := \hat{b}_j^{\dagger} \hat{b}_j$, and $\mathcal{J}_0$ is the Bessel function of the first kind. For a typical shaking frequency $\omega = 2\pi \times 18\,$kHz and amplitude $\mathcal{A} = 5\,$nm, $F_0 d = 5\,$kHz, $\eta = 0.3$, $\Omega = 0.75\,$kHz, $U_{ab}/(\hbar\omega) = 0.1$, and $\Omega/(\hbar\omega) = 0.04$. So the effect of the corrections coming from $\hat{H}_1$ and $\hat{H}_2$ is small in the experiment. We calculate an effective phase diagram (Figs.~2 and \ref{figs:phase_diagram}) by minimizing $\hat{H}_{\text{eff}}$ for system sizes $L=16,32,64,128,256$ with unity filling using two-site DMRG with open boundary conditions, singular value cutoff $10^{-8}$, maximum bond dimension 200, and up to 4 particles per site.
	
\section{Phase diagram and critical points}

\begin{figure}[h]
	\includegraphics[width=1\linewidth]{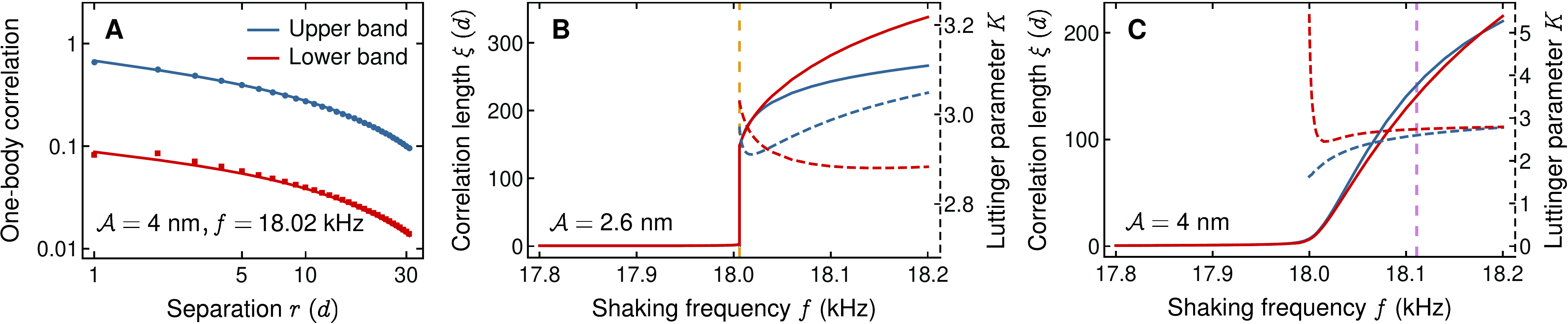}
	\centering
	\caption{\textbf{Single-particle correlations.} (\textbf{A}) Average correlations $\langle\langle \hat{a}_i^{\dagger} \hat{a}_{i+r} \rangle\rangle$ and $|\langle\langle \hat{b}_i^{\dagger} \hat{b}_{i+r} \rangle\rangle|$ in the ground state with $L=64$ sites, fitted with an exponential times a power law. (\textbf{B-C}) Correlation length (solid lines) and Luttinger parameter (dashed lines) extracted from the fits across first-order and continuous transitions denoted by vertical lines (cf. Fig.~\ref{figs:entanglement_central_charge}).}
	\label{figs:sp_corrlation}
\end{figure}

To characterize the ground states of $\hat{H}_{\text{eff}}$, we find the single-particle correlations in the bulk, e.g., $\langle\langle \hat{b}_i^{\dagger} \hat{b}_{i+r} \rangle\rangle$, by discarding $L/4$ sites from either end and averaging over all remaining pairs of sites with separation $r$ (double brackets in this section denote this bulk averaging). We find these correlations are well fitted by $|\langle\langle \hat{b}_i^{\dagger} \hat{b}_{i+r} \rangle\rangle| \approx C r^{-1/(2K)} e^{-r/\xi}$ (and similarly for $\langle\langle \hat{a}_i^{\dagger} \hat{a}_{i+r} \rangle\rangle$), where $\xi$ is the correlation length and $K$ is an effective Luttinger parameter, as shown in Fig.~\ref{figs:sp_corrlation}A. In the Mott phase, $\xi$ is finite and the correlations decay exponentially at large distances. In the superfluid phase, $\xi$ becomes much larger than $L$ and the correlations follow a power law set by the drive parameters. At weak shaking amplitudes, both $\xi$ and $K$ jump discontinuously across a first-order transition (Fig.~\ref{figs:sp_corrlation}B), whereas at large amplitudes, they vary continuously across a Kosterlitz-Thouless (KT) transition (Fig.~\ref{figs:sp_corrlation}C). For very strong shaking ($\mathcal{A}\to\infty$), the KT transition occurs in a single hybridized band at $K = 2$ \cite{sRachel2012}.

To determine the phase boundary more robustly, we analyze the bipartite entanglement in the ground state which shows universal scaling close to a critical point where the system maps onto a conformal field theory (CFT) \cite{sCalabrese2004}. In particular, the von Neumann entanglement entropy for a bipartition at site $i$ is given by
\begin{equation}
    S_{\text{vN}} = \frac{c}{6} \log \left[ \frac{L}{\pi} \sin \left( \frac{\pi i}{L} \right) \right] + c_1\;,
    \label{eqs:CFTentropy}
\end{equation}
where $c_1$ is a nonuniversal constant and $c$ is called a central charge which determines how strongly the entanglement entropy varies near $i=L/2$. 
\begin{figure}[h]
	\includegraphics[width=1\linewidth]{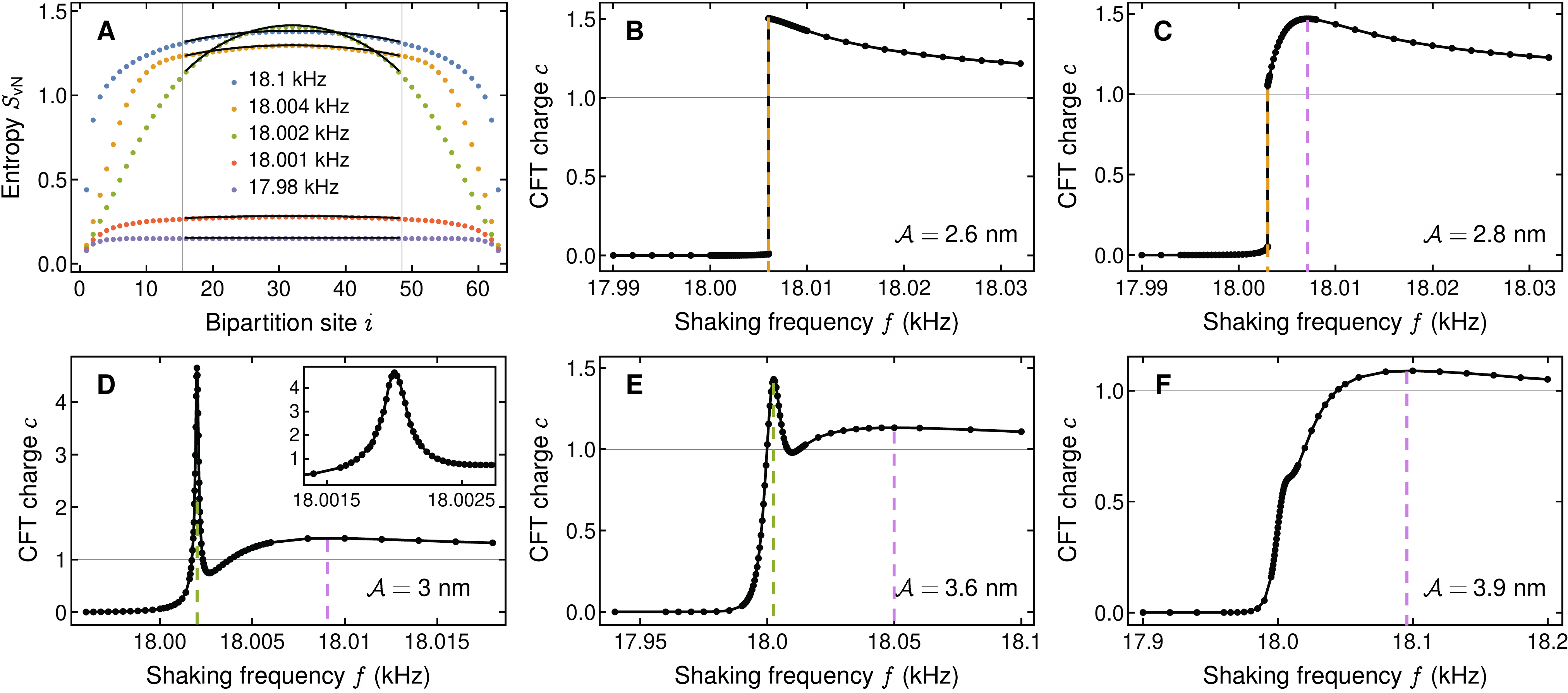}
	\centering
	\caption{\textbf{Entanglement and central charge.} (\textbf{A}) Von Neumann entanglement entropy across bipartitions of the ground state for $L=64$ and $\mathcal{A} = 3$~nm with different shaking frequencies, fitted with the conformal field theory (CFT) prediction in Eq.~\eqref{eqs:CFTentropy}. (\textbf{B-F}) Fitted central charge $c$ as a function of the shaking frequency at increasing shaking amplitudes, exhibiting discontinuous jumps and smooth peaks characteristic of discontinuous (first-order) and continuous phase transitions, respectively (shown by the vertical dashed lines, with the same color convention as in Fig.~\ref{figs:phase_diagram}).}
	\label{figs:entanglement_central_charge}
\end{figure}
As shown in Fig.~\ref{figs:entanglement_central_charge}A, the bulk entropy variation (for $L=64$) is very well approximated by Eq.~\eqref{eqs:CFTentropy} with $c$ and $c_1$ as fit parameters. The fitted charge $c$ is 0 well inside the Mott phase and 1 well inside the superfluid phase. For weak amplitudes (Fig.~\ref{figs:entanglement_central_charge}B), this change occurs discontinuously across a first-order transition where $c$ jumps from 0 to its maximum value, coinciding with the jump in the correlation lengths in Fig.~\ref{figs:sp_corrlation}B. On the other hand, the large-amplitude variation in Fig.~\ref{figs:entanglement_central_charge}F is continuous and exhibits a smooth peak, similar to what happens for a Mott-superfluid transition in a single band \cite{sEjima2012}. As argued in Ref.~\cite{sEjima2012}, the critical point can be traced to the location of this maximum. For intermediate amplitudes, the behavior is more complicated. In Fig.~\ref{figs:entanglement_central_charge}C, one finds both a jump and a smooth peak, signaling two back-to-back phase transitions. As the amplitude is increased, the first-order jump changes into a sharp peak (Fig.~\ref{figs:entanglement_central_charge}D) which then diminishes and eventually disappears [Figs.~\ref{figs:entanglement_central_charge}(E-F)]. Tracing these peaks and jumps leads to the phase diagram in Fig.~\ref{figs:phase_diagram}A. 
\begin{figure}[h]
	\includegraphics[width=0.72\linewidth]{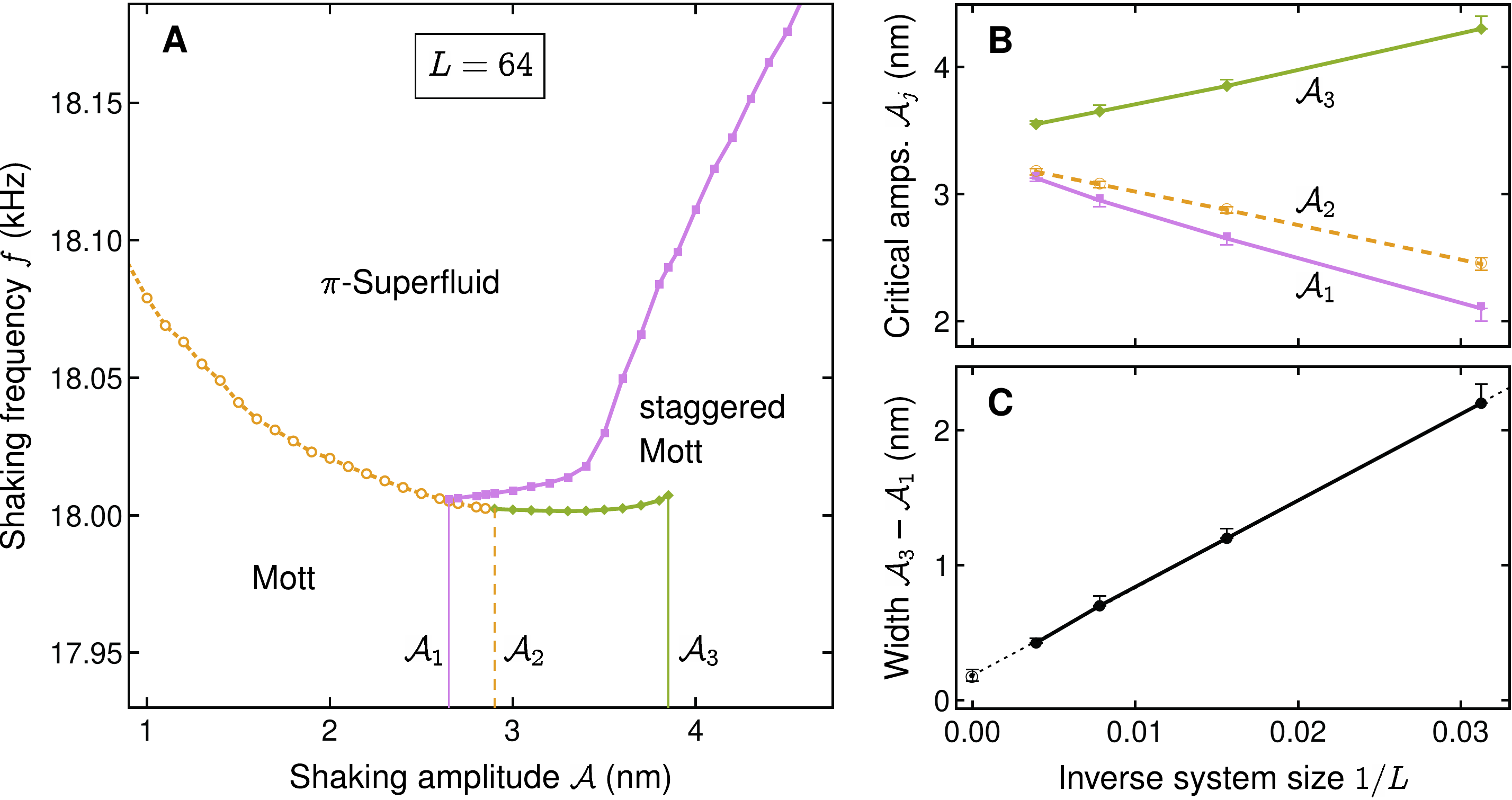}
	\centering
	\caption{\textbf{Phase boundaries. (\textbf{A})} Phase diagram obtained by tracing the location of the jumps and peaks in the fitted CFT charge in Fig.~\ref{figs:entanglement_central_charge}. The Mott to $\pi$-superfluid transition is discontinuous (first-order) for amplitudes $\mathcal{A} < \mathcal{A}_1$ (dashed yellow) and continuous for $\mathcal{A} > \mathcal{A}_1$ (solid blue). There is also a transition between non-staggered Mott and staggered Mott (alternating phases) for $\mathcal{A}_1 < \mathcal{A} < \mathcal{A}_3$ that is discontinuous for $\mathcal{A} < \mathcal{A}_2$ and continuous (solid green) for $\mathcal{A} > \mathcal{A}_2$. (\textbf{B-C}) The interval $\mathcal{A}_1$-$\mathcal{A}_3$ shrinks as $1/L$, indicating the non-staggered to staggered Mott transition could be a finite-size effect.}
	\label{figs:phase_diagram}
\end{figure}
Between amplitudes $\mathcal{A}_1$ and $\mathcal{A}_3$, we find two closely-spaced phase transitions. The intermediate phase can be identified as a staggered Mott phase where the phase correlations alternate in sign, $\langle\langle \hat{a}_i^{\dagger} \hat{a}_{i+r} \rangle\rangle \sim (-1)^r e^{-r/\xi}$. 
\begin{figure}[h]
	\includegraphics[width=1\linewidth]{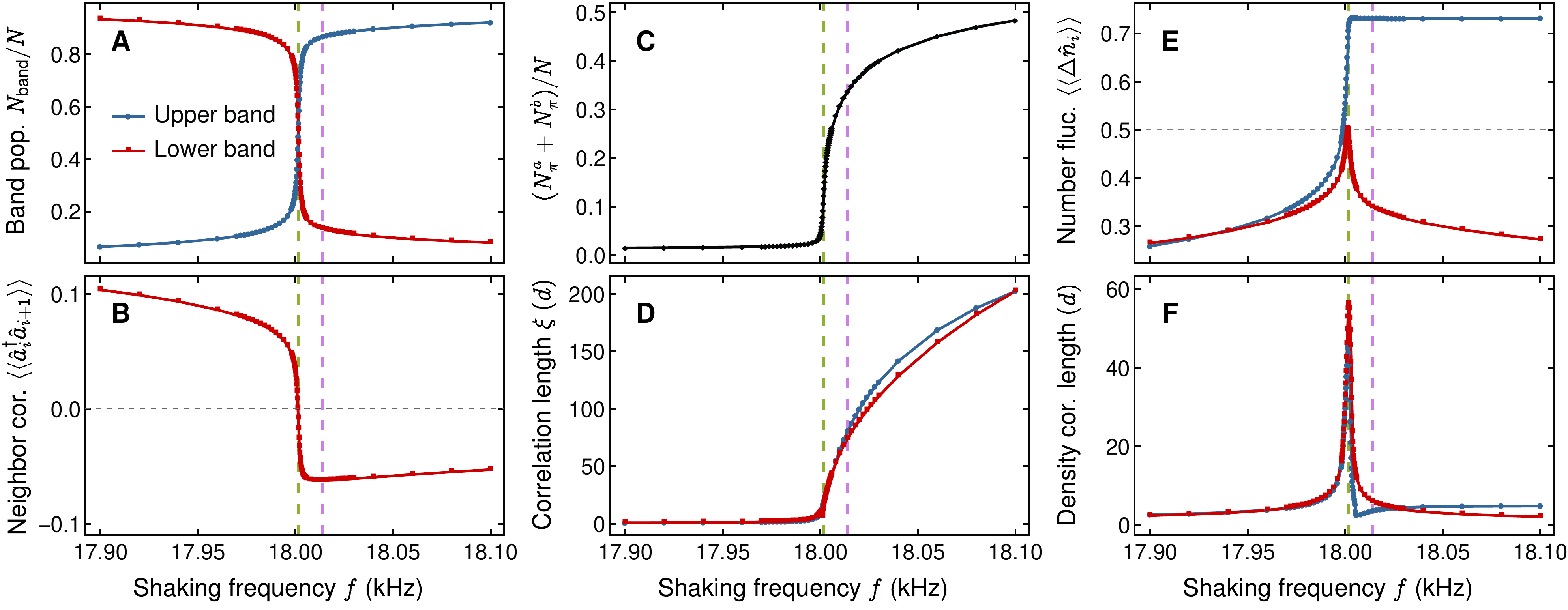}
	\centering
	\caption{\textbf{Evolution of ground state across phase transitions}. Ground-state properties for $L=64$ and $\mathcal{A} = 3.3$~nm as a function of the shaking frequency across back-to-back transitions from a non-staggered to a staggered Mott and from a staggered Mott to a $\pi$-superfluid (Fig.~\ref{figs:phase_diagram}A). At the former, (\textbf{A}) band populations are inverted, (\textbf{B}) the nearest-neighbor correlation in the lowest band flips sign, (\textbf{C}) populations at the band edge and (\textbf{D}) single-particle correlation lengths start growing rapidly, (\textbf{E}) number fluctuations are maximized, and (\textbf{F}) density-density correlation length diverges (for $\mathcal{A}_2 < \mathcal{A} < \mathcal{A}_3$). The density correlation length is found by estimating where the correlation $\langle\langle\hat{n}_i, \hat{n}_{i+r}\rangle\rangle$ decays to a hundredth of its maximum value, where $ \langle \hat{n}_i, \hat{n}_j \rangle := \langle \hat{n}_i \hat{n}_j \rangle - \langle \hat{n}_i \rangle \langle \hat{n}_j \rangle $.}
	\label{figs:observable}
\end{figure}
At the non-staggered to staggered Mott transition, the band occupations are inverted (Fig.~\ref{figs:observable}A), the nearest-neighbor correlation $\langle\langle \hat{a}_i^{\dagger} \hat{a}_{i+1} \rangle\rangle$ flips sign (Fig.~\ref{figs:observable}B), and the density fluctuation in the lowest band reaches a maximum (Fig.~\ref{figs:observable}E). For $\mathcal{A}_2 < \mathcal{A} < \mathcal{A}_3$, this transition is continuous and accompanied by a sharp peak in the density-density correlation length at the critical point (Fig.~\ref{figs:observable}F). Above $\mathcal{A}_3$, this transition between non-staggered and staggered Mott insulators becomes a crossover. Furthermore, the `phase transition' between different Mott states could be a finite-size effect, as the interval $\mathcal{A}_1$-$\mathcal{A}_3$ shrinks as $1/L$, see Figs.~\ref{figs:phase_diagram}(B-C). This points to a simpler picture in the thermodynamic limit, where one only has a first-order non-staggered Mott to $\pi$-superfluid transition below a critical amplitude $\mathcal{A}^* \approx 3.3$ nm and a continuous staggered Mott to $\pi$-superfluid transition for $\mathcal{A} > \mathcal{A}^*$.

\section{\label{sec:3band}Three-band simulations}

So far we have considered only the lowest two bands and ignored the coupling to higher bands. In this section we include the third band and show that it has a significant effect on the dynamics only for large-amplitude sweeps over long periods of time. We extend the modeling in Sec.~\ref{sec:model} by writing the field operator as
\begin{equation}
\hat{\psi} (y) = 
\sum_{j} w_1(y-y_j)\push \hat{a}_j + w_2 (y-y_j)\push \hat{b}_j + w_3 (y-y_j)\push \hat{c}_j \push,
\end{equation}
where $\hat{c}_j$ annihilates a particle at site $y_j$ in the third (second exited) band and $w_3$ is the corresponding Wannier function. Substituting this expansion into Eq.~\eqref{suppeq:latframehamil} and retaining the most significant terms for a deep lattice, we find
\begin{equation}
    \hat{H}(t) = \hat{T} + \hat{U} + \hat{S}(t) \push,
\end{equation}
where
\begin{align}
    \hat{T} &:= \sum_j (\varepsilon_a - \varepsilon_b) \push \hat{n}^a_j + (\varepsilon_c - \varepsilon_b) \push \hat{n}^c_j
    - J_a \sum_{\langle i,j \rangle} \hat{a}_i^{\dagger} \hat{a}_j
    + J_b \sum_{\langle i,j \rangle} \hat{b}_i^{\dagger} \hat{b}_j
    + J_b^{\prime} \sum_{\langle\langle i,j \rangle\rangle} \hat{b}_i^{\dagger} \hat{b}_j
    + \sum_{r=1}^{r_{\text{max}}} J_c(r) \big(\hat{c}_i^{\dagger} \hat{c}_{i+r} + \text{h.c.}\big) \push, \\
    \nonumber \hat{U} &:= \sum_j 
    \frac{U_a}{2} \hat{a}_j^{\dagger} \hat{a}_j^{\dagger} \hat{a}_j \hat{a}_j 
    + \frac{U_b}{2} \hat{b}_j^{\dagger} \hat{b}_j^{\dagger} \hat{b}_j \hat{b}_j 
    + \frac{U_c}{2} \hat{c}_j^{\dagger} \hat{c}_j^{\dagger} \hat{c}_j \hat{c}_j
    + U_{ab} \hat{n}^a_j \hat{n}^b_j
    + U_{bc} \hat{n}^b_j \hat{n}^c_j
    + U_{ac} \hat{n}^a_j \hat{n}^c_j
    + U^{ac}_{bb} \big( \hat{a}_j^{\dagger} \hat{c}_j^{\dagger} \hat{b}_j \hat{b}_j + \text{h.c.} \big) \\
    & \hspace{1cm} + \bigg( \frac{U_{ab}}{4} \hat{a}_j^{\dagger} \hat{a}_j^{\dagger} \hat{b}_j \hat{b}_j
    + \frac{U_{bc}}{4} \hat{b}_j^{\dagger} \hat{b}_j^{\dagger} \hat{c}_j \hat{c}_j
    + \frac{U_{ac}}{4} \hat{a}_j^{\dagger} \hat{a}_j^{\dagger} \hat{c}_j \hat{c}_j
    + U^{ac}_{aa} \hat{a}_j^{\dagger} \hat{n}^a_j \hat{c}_j
    + U^{ac}_{cc} \hat{a}_j^{\dagger} \hat{n}^c_j \hat{c}_j
    + 2 U^{ac}_{bb} \hat{a}_j^{\dagger} \hat{n}^b_j \hat{c}_j
    + \text{h.c.} \pull \bigg) \push, \\[1.5ex]
    \hspace{-10cm}\hat{S}(t) &:= F(t) d \sum_j j \big(\hat{n}^a_j + \hat{n}^b_j + \hat{n}^c_j\big)
    + \alpha_{ab} \big(\hat{a}_j^{\dagger} \hat{b}_j + \hat{b}_j^{\dagger} \hat{a}_j\big)
    + \alpha_{bc} \big(\hat{b}_j^{\dagger} \hat{c}_j + \hat{c}_j^{\dagger} \hat{b}_j\big) \push,
\end{align}
where $\varepsilon_c$ is the average energy of the third band, $\hat{n}^x_j \equiv \hat{x}_j^{\dagger} \hat{x}_j$ denotes the on-site occupation in each band, and
\begin{equation}
\alpha_{bc} = \frac{1}{d}\int \pull dy \push y \push w_2(y) w_3(y) \;
\end{equation}
gives the coupling between the second and the third band. Note there is no direct coupling between the first and the third band as those Wannier functions have the same parity and the shaking represents a linear potential in the lattice frame. For the experimental lattice depth $V_0 = 8.4 E_r$, the tunneling amplitudes $J_c(r)$ fall off rapidly with separation $r$ and we take $r_{\text{max}} = 3$ in the simulations which accurately reproduces the band structure. As in Sec.~\ref{sec:model}, we find the parameters $\varepsilon_b - \varepsilon_a = 19.7\,$kHz, $\varepsilon_c - \varepsilon_b = 17.1$~kHz, $J_a = 0.12\,$kHz, $J_b = 1.29\,$kHz, $J_b^{\prime} = 0.17\,$kHz, $J_c(1) = -3.71\,$kHz, $J_c(2) = 0.45\,$kHz, $J_c(3) = -0.31\,$kHz, $U_a = 2.88\,$kHz, $U_b = 1.73\,$kHz, $U_c = 1.22\,$kHz, $U_{ab} = 2.36\,$kHz, $U_{bc} = 1.90\,$kHz, $U_{ac} = 1.49\,$kHz, $U^{ac}_{bb} = -0.38\,$kHz, $U^{ac}_{aa} = 0.91\,$kHz, $U^{ac}_{cc} = -0.12\,$kHz, $\alpha_{ab} = 0.15$, and $\alpha_{bc} = -0.23$. We simulate the experimental sweeps using adaptive tDMRG with the same numerical parameters as in Sec.~\ref{sec:model}.

\begin{figure}[h]
	\includegraphics[width=1\linewidth]{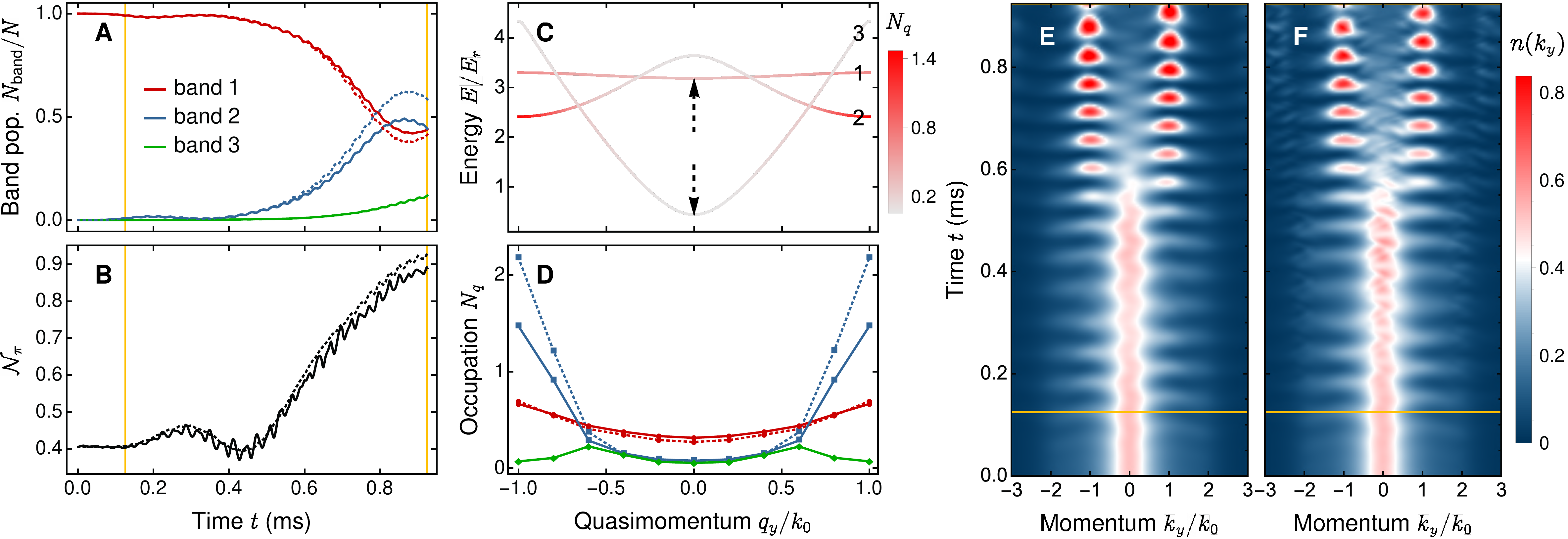}
	\centering
	\caption{\textbf{Simulation of an experimental sweep.} tDMRG simulation of a two-step sweep sequence with $L=10$, including the lowest two bands (dotted lines) and the lowest three bands (solid lines). First, the shaking amplitude $\mathcal{A}$ is ramped linearly from 0 to 5.8 nm over 0.125 ms (yellow grid line) with shaking frequency $f = 15\,$kHz. Next, $f$ is swept linearly from 15 kHz to 21 kHz over $\tau = 0.8$~ms. (\textbf{A}) Relative population of the bare bands, showing a small third-band excitation, and (\textbf{B}) the resulting population $\mathcal{N}_{\pi}$. (\textbf{C}) Occupation of bare bands in the rotating frame at the end of the sweep, where the first band is raised by $h f$ and the third band is lowered by $h f$. Black arrows indicate the relative movement of these band during the sweep, highlighting that the third band is always off-resonant at the edge of the Brillouin zone. (\textbf{D}) Final quasimomentum occupations showing sharp peaks at $\pm \hbar k_0$ and small peaks in the third band where it crosses the second band. These give rise to satellite peaks in the normalized momentum distribution (\textbf{F}) absent in a two-band simulation (\textbf{E}).}
	\label{figs:3band}
\end{figure}

Figure~\ref{figs:3band} shows the evolution during such a sweep at relatively strong shaking amplitude ($\mathcal{A} = 5.8$ nm) where the shaking frequency $f$ is swept through resonance from 15 kHz to 21 kHz over 0.8 ms. Only about $10\%$ of the particles are excited to the third band toward the end of the sweep (Fig.~\ref{figs:3band}A). As shown in Figs.~\ref{figs:3band}(C-D), this excitation occurs primarily at quasimomenta where the second and third bands are resonant. Note the particles migrate from the first to the second band near the edge of the Brillouin zone, which remains off-resonant with the third band throughout the sweep. The small third-band population leads to satellite peaks in the plane wave basis (Fig.~\ref{figs:3band}F), as observed in the experiment (Fig.~\ref{figs2}A) but absent in a two-band simulation (Fig.~\ref{figs:3band}E). Thus, the observable $\mathcal{N}_{\pi}$ is slightly reduced by the presence of the third band (Fig.~\ref{figs:3band}B). It also exhibits stronger oscillations at frequency $2f$. This frequency doubling arises because $n_{\pi}$ is calculated by adding counter-propagating modes (see Fig.~\ref{figs2}A).

\begin{figure}[h]
	\includegraphics[width=1\linewidth]{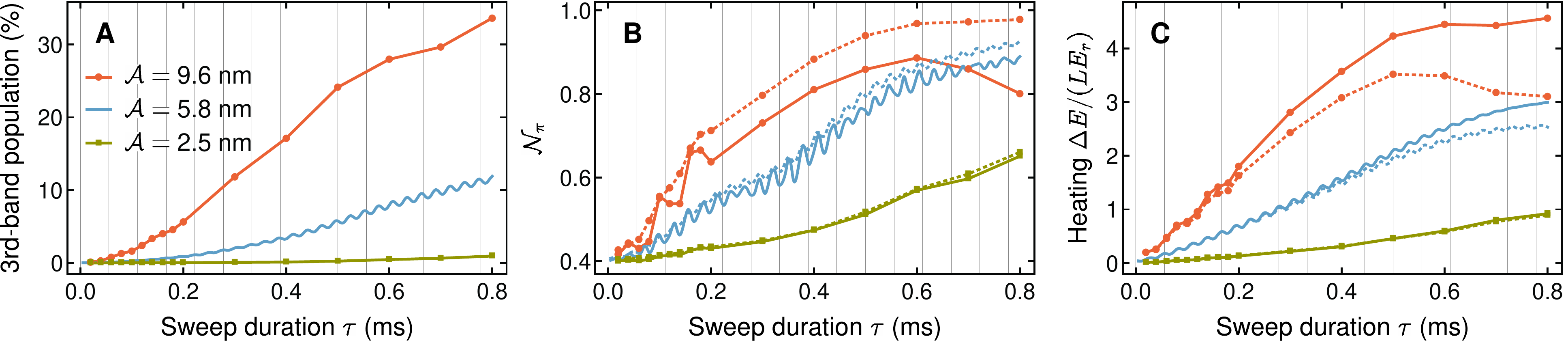}
	\centering
	\caption{\textbf{Three-band vs two-band simulations}. Occupations and energies at the end of frequency sweeps with different amplitudes $\mathcal{A}$ and sweep durations $\tau$, from two-band (dotted) and three-band (solid) simulations. (\textbf{A}) Population of the third band grows with both $\mathcal{A}$ and $\tau$. (\textbf{B}) $\mathcal{N}_{\pi}$ is reduced and oscillates more strongly due to population transfer to the third band. (\textbf{C}) Third-band excitation causes more heating (relative to the undriven Hamiltonian). Vertical lines are spaced by the average frequency $\bar{f} = 18$~kHz, showing that the fast oscillations occur at frequency $2\bar{f}$ due to micromotion (see Sec.~\ref{sec:micromotion}).}
	\label{figs:3bandvs2band}
\end{figure}

Figure~\ref{figs:3bandvs2band} shows that the third band is more significant at larger shaking amplitudes and longer sweep durations. In particular, its population is less than $10\%$ for amplitudes $\mathcal{A} < 5\,$nm and practically zero for $\mathcal{A} < 2.6\,$nm (Fig.~\ref{figs:3bandvs2band}A) where the phase transition is discontinuous (Fig.~\ref{figs:phase_diagram}A). Stronger shaking over a long period allows more particles to be transferred from the second to the third band, decreasing $\mathcal{N}_{\pi}$ (Fig.~\ref{figs:3bandvs2band}B) and causing more heating (Fig.~\ref{figs:3bandvs2band}C). Note that the energy is bounded for a finite number of bands, so the heating eventually saturates as longer sweeps become adiabatic. Figure~\ref{figs:3bandvs2band}B shows that as a function of the sweep duration, the final value of $\mathcal{N}_{\pi}$ exhibits fast oscillations at twice the average shaking frequency (for linear ramps from 15 to 21 kHz), which originates from the fast oscillations in its time evolution (Fig.~\ref{figs:3band}B).


\begin{thebibliography}{39}%
\makeatletter
\providecommand \@ifxundefined [1]{%
 \@ifx{#1\undefined}
}%
\providecommand \@ifnum [1]{%
 \ifnum #1\expandafter \@firstoftwo
 \else \expandafter \@secondoftwo
 \fi
}%
\providecommand \@ifx [1]{%
 \ifx #1\expandafter \@firstoftwo
 \else \expandafter \@secondoftwo
 \fi
}%
\providecommand \natexlab [1]{#1}%
\providecommand \enquote  [1]{``#1''}%
\providecommand \bibnamefont  [1]{#1}%
\providecommand \bibfnamefont [1]{#1}%
\providecommand \citenamefont [1]{#1}%
\providecommand \href@noop [0]{\@secondoftwo}%
\providecommand \href [0]{\begingroup \@sanitize@url \@href}%
\providecommand \@href[1]{\@@startlink{#1}\@@href}%
\providecommand \@@href[1]{\endgroup#1\@@endlink}%
\providecommand \@sanitize@url [0]{\catcode `\\12\catcode `\$12\catcode
  `\&12\catcode `\#12\catcode `\^12\catcode `\_12\catcode `\%12\relax}%
\providecommand \@@startlink[1]{}%
\providecommand \@@endlink[0]{}%
\providecommand \url  [0]{\begingroup\@sanitize@url \@url }%
\providecommand \@url [1]{\endgroup\@href {#1}{\urlprefix }}%
\providecommand \urlprefix  [0]{URL }%
\providecommand \Eprint [0]{\href }%
\providecommand \doibase [0]{https://doi.org/}%
\providecommand \selectlanguage [0]{\@gobble}%
\providecommand \bibinfo  [0]{\@secondoftwo}%
\providecommand \bibfield  [0]{\@secondoftwo}%
\providecommand \translation [1]{[#1]}%
\providecommand \BibitemOpen [0]{}%
\providecommand \bibitemStop [0]{}%
\providecommand \bibitemNoStop [0]{.\EOS\space}%
\providecommand \EOS [0]{\spacefactor3000\relax}%
\providecommand \BibitemShut  [1]{\csname bibitem#1\endcsname}%
\let\auto@bib@innerbib\@empty
\bibitem [{\citenamefont {Kibble}(1980)}]{kibble1980some}%
  \BibitemOpen
  \bibfield  {author} {\bibinfo {author} {\bibfnamefont {T.~W.}\ \bibnamefont
  {Kibble}},\ }\href {https://doi.org/10.1016/0370-1573(80)90091-5} {\bibfield
  {journal} {\bibinfo  {journal} {Physics Reports}\ }\textbf {\bibinfo {volume}
  {67}},\ \bibinfo {pages} {183} (\bibinfo {year} {1980})}\BibitemShut
  {NoStop}%
\bibitem [{\citenamefont {McLerran}(1986)}]{mclerran1986physics}%
  \BibitemOpen
  \bibfield  {author} {\bibinfo {author} {\bibfnamefont {L.}~\bibnamefont
  {McLerran}},\ }\href {https://doi.org/10.1103/RevModPhys.58.1021} {\bibfield
  {journal} {\bibinfo  {journal} {Reviews of Modern Physics}\ }\textbf
  {\bibinfo {volume} {58}},\ \bibinfo {pages} {1021} (\bibinfo {year}
  {1986})}\BibitemShut {NoStop}%
\bibitem [{\citenamefont {Sachdev}(2011)}]{sachdev2011}%
  \BibitemOpen
  \bibfield  {author} {\bibinfo {author} {\bibfnamefont {S.}~\bibnamefont
  {Sachdev}},\ }\href {https://doi.org/10.1017/CBO9780511973765} {\emph
  {\bibinfo {title} {Quantum Phase Transitions}}},\ \bibinfo {edition} {2nd}\
  ed.\ (\bibinfo  {publisher} {Cambridge University Press},\ \bibinfo {year}
  {2011})\BibitemShut {NoStop}%
\bibitem [{\citenamefont {Lifshitz}\ and\ \citenamefont
  {Kagan}(1972)}]{lifshitz1972quantum}%
  \BibitemOpen
  \bibfield  {author} {\bibinfo {author} {\bibfnamefont {I.~M.}\ \bibnamefont
  {Lifshitz}}\ and\ \bibinfo {author} {\bibfnamefont {Y.}~\bibnamefont
  {Kagan}},\ }\href {http://www.jetp.ac.ru/cgi-bin/e/index/e/35/1/p206?a=list}
  {\bibfield  {journal} {\bibinfo  {journal} {Soviet Journal of Experimental
  and Theoretical Physics}\ }\textbf {\bibinfo {volume} {35}},\ \bibinfo
  {pages} {206} (\bibinfo {year} {1972})}\BibitemShut {NoStop}%
\bibitem [{\citenamefont {Coleman}(1977)}]{coleman1977fate}%
  \BibitemOpen
  \bibfield  {author} {\bibinfo {author} {\bibfnamefont {S.}~\bibnamefont
  {Coleman}},\ }\href {https://doi.org/10.1103/PhysRevD.15.2929} {\bibfield
  {journal} {\bibinfo  {journal} {Physical Review D}\ }\textbf {\bibinfo
  {volume} {15}},\ \bibinfo {pages} {2929} (\bibinfo {year}
  {1977})}\BibitemShut {NoStop}%
\bibitem [{\citenamefont {Vilenkin}(1983)}]{vilenkin1983birth}%
  \BibitemOpen
  \bibfield  {author} {\bibinfo {author} {\bibfnamefont {A.}~\bibnamefont
  {Vilenkin}},\ }\href {https://doi.org/10.1103/PhysRevD.27.2848} {\bibfield
  {journal} {\bibinfo  {journal} {Physical Review D}\ }\textbf {\bibinfo
  {volume} {27}},\ \bibinfo {pages} {2848} (\bibinfo {year}
  {1983})}\BibitemShut {NoStop}%
\bibitem [{\citenamefont {Fialko}\ \emph {et~al.}(2015)\citenamefont {Fialko},
  \citenamefont {Opanchuk}, \citenamefont {Sidorov}, \citenamefont {Drummond},\
  and\ \citenamefont {Brand}}]{fialko2015fate}%
  \BibitemOpen
  \bibfield  {author} {\bibinfo {author} {\bibfnamefont {O.}~\bibnamefont
  {Fialko}}, \bibinfo {author} {\bibfnamefont {B.}~\bibnamefont {Opanchuk}},
  \bibinfo {author} {\bibfnamefont {A.}~\bibnamefont {Sidorov}}, \bibinfo
  {author} {\bibfnamefont {P.}~\bibnamefont {Drummond}},\ and\ \bibinfo
  {author} {\bibfnamefont {J.}~\bibnamefont {Brand}},\ }\href
  {https://doi.org/10.1209/0295-5075/110/56001} {\bibfield  {journal} {\bibinfo
   {journal} {EPL (Europhysics Letters)}\ }\textbf {\bibinfo {volume} {110}},\
  \bibinfo {pages} {56001} (\bibinfo {year} {2015})}\BibitemShut {NoStop}%
\bibitem [{\citenamefont {Fialko}\ \emph {et~al.}(2017)\citenamefont {Fialko},
  \citenamefont {Opanchuk}, \citenamefont {Sidorov}, \citenamefont {Drummond},\
  and\ \citenamefont {Brand}}]{fialko2017universe}%
  \BibitemOpen
  \bibfield  {author} {\bibinfo {author} {\bibfnamefont {O.}~\bibnamefont
  {Fialko}}, \bibinfo {author} {\bibfnamefont {B.}~\bibnamefont {Opanchuk}},
  \bibinfo {author} {\bibfnamefont {A.~I.}\ \bibnamefont {Sidorov}}, \bibinfo
  {author} {\bibfnamefont {P.~D.}\ \bibnamefont {Drummond}},\ and\ \bibinfo
  {author} {\bibfnamefont {J.}~\bibnamefont {Brand}},\ }\href
  {https://doi.org/10.1088/1361-6455/50/2/024003} {\bibfield  {journal}
  {\bibinfo  {journal} {Journal of Physics B: Atomic, Molecular and Optical
  Physics}\ }\textbf {\bibinfo {volume} {50}},\ \bibinfo {pages} {024003}
  (\bibinfo {year} {2017})}\BibitemShut {NoStop}%
\bibitem [{\citenamefont {Ng}\ \emph {et~al.}(2021)\citenamefont {Ng},
  \citenamefont {Opanchuk}, \citenamefont {Thenabadu}, \citenamefont {Reid},\
  and\ \citenamefont {Drummond}}]{ng2021fate}%
  \BibitemOpen
  \bibfield  {author} {\bibinfo {author} {\bibfnamefont {K.~L.}\ \bibnamefont
  {Ng}}, \bibinfo {author} {\bibfnamefont {B.}~\bibnamefont {Opanchuk}},
  \bibinfo {author} {\bibfnamefont {M.}~\bibnamefont {Thenabadu}}, \bibinfo
  {author} {\bibfnamefont {M.}~\bibnamefont {Reid}},\ and\ \bibinfo {author}
  {\bibfnamefont {P.~D.}\ \bibnamefont {Drummond}},\ }\href
  {https://doi.org/10.1103/PRXQuantum.2.010350} {\bibfield  {journal} {\bibinfo
   {journal} {PRX Quantum}\ }\textbf {\bibinfo {volume} {2}},\ \bibinfo {pages}
  {010350} (\bibinfo {year} {2021})}\BibitemShut {NoStop}%
\bibitem [{\citenamefont {Greiner}\ \emph {et~al.}(2002)\citenamefont
  {Greiner}, \citenamefont {Mandel}, \citenamefont {Esslinger}, \citenamefont
  {H{\"a}nsch},\ and\ \citenamefont {Bloch}}]{greiner2002quantum}%
  \BibitemOpen
  \bibfield  {author} {\bibinfo {author} {\bibfnamefont {M.}~\bibnamefont
  {Greiner}}, \bibinfo {author} {\bibfnamefont {O.}~\bibnamefont {Mandel}},
  \bibinfo {author} {\bibfnamefont {T.}~\bibnamefont {Esslinger}}, \bibinfo
  {author} {\bibfnamefont {T.~W.}\ \bibnamefont {H{\"a}nsch}},\ and\ \bibinfo
  {author} {\bibfnamefont {I.}~\bibnamefont {Bloch}},\ }\href
  {https://doi.org/10.1038/415039a} {\bibfield  {journal} {\bibinfo  {journal}
  {Nature}\ }\textbf {\bibinfo {volume} {415}},\ \bibinfo {pages} {39}
  (\bibinfo {year} {2002})}\BibitemShut {NoStop}%
\bibitem [{\citenamefont {Gross}\ and\ \citenamefont
  {Bloch}(2017)}]{gross2017quantum}%
  \BibitemOpen
  \bibfield  {author} {\bibinfo {author} {\bibfnamefont {C.}~\bibnamefont
  {Gross}}\ and\ \bibinfo {author} {\bibfnamefont {I.}~\bibnamefont {Bloch}},\
  }\href {https://doi.org/10.1126/science.aal3837} {\bibfield  {journal}
  {\bibinfo  {journal} {Science}\ }\textbf {\bibinfo {volume} {357}},\ \bibinfo
  {pages} {995} (\bibinfo {year} {2017})}\BibitemShut {NoStop}%
\bibitem [{\citenamefont {Struck}\ \emph {et~al.}(2013)\citenamefont {Struck},
  \citenamefont {Weinberg}, \citenamefont {{\"O}lschl{\"a}ger}, \citenamefont
  {Windpassinger}, \citenamefont {Simonet}, \citenamefont {Sengstock},
  \citenamefont {H{\"o}ppner}, \citenamefont {Hauke}, \citenamefont {Eckardt},
  \citenamefont {Lewenstein} \emph {et~al.}}]{struck2013engineering}%
  \BibitemOpen
  \bibfield  {author} {\bibinfo {author} {\bibfnamefont {J.}~\bibnamefont
  {Struck}}, \bibinfo {author} {\bibfnamefont {M.}~\bibnamefont {Weinberg}},
  \bibinfo {author} {\bibfnamefont {C.}~\bibnamefont {{\"O}lschl{\"a}ger}},
  \bibinfo {author} {\bibfnamefont {P.}~\bibnamefont {Windpassinger}}, \bibinfo
  {author} {\bibfnamefont {J.}~\bibnamefont {Simonet}}, \bibinfo {author}
  {\bibfnamefont {K.}~\bibnamefont {Sengstock}}, \bibinfo {author}
  {\bibfnamefont {R.}~\bibnamefont {H{\"o}ppner}}, \bibinfo {author}
  {\bibfnamefont {P.}~\bibnamefont {Hauke}}, \bibinfo {author} {\bibfnamefont
  {A.}~\bibnamefont {Eckardt}}, \bibinfo {author} {\bibfnamefont
  {M.}~\bibnamefont {Lewenstein}}, \emph {et~al.},\ }\href
  {https://doi.org/10.1038/nphys2750} {\bibfield  {journal} {\bibinfo
  {journal} {Nature Physics}\ }\textbf {\bibinfo {volume} {9}},\ \bibinfo
  {pages} {738} (\bibinfo {year} {2013})}\BibitemShut {NoStop}%
\bibitem [{\citenamefont {Trenkwalder}\ \emph {et~al.}(2016)\citenamefont
  {Trenkwalder}, \citenamefont {Spagnolli}, \citenamefont {Semeghini},
  \citenamefont {Coop}, \citenamefont {Landini}, \citenamefont {Castilho},
  \citenamefont {Pezze}, \citenamefont {Modugno}, \citenamefont {Inguscio},
  \citenamefont {Smerzi} \emph {et~al.}}]{trenkwalder2016quantum}%
  \BibitemOpen
  \bibfield  {author} {\bibinfo {author} {\bibfnamefont {A.}~\bibnamefont
  {Trenkwalder}}, \bibinfo {author} {\bibfnamefont {G.}~\bibnamefont
  {Spagnolli}}, \bibinfo {author} {\bibfnamefont {G.}~\bibnamefont
  {Semeghini}}, \bibinfo {author} {\bibfnamefont {S.}~\bibnamefont {Coop}},
  \bibinfo {author} {\bibfnamefont {M.}~\bibnamefont {Landini}}, \bibinfo
  {author} {\bibfnamefont {P.}~\bibnamefont {Castilho}}, \bibinfo {author}
  {\bibfnamefont {L.}~\bibnamefont {Pezze}}, \bibinfo {author} {\bibfnamefont
  {G.}~\bibnamefont {Modugno}}, \bibinfo {author} {\bibfnamefont
  {M.}~\bibnamefont {Inguscio}}, \bibinfo {author} {\bibfnamefont
  {A.}~\bibnamefont {Smerzi}}, \emph {et~al.},\ }\href
  {https://doi.org/10.1038/nphys3743} {\bibfield  {journal} {\bibinfo
  {journal} {Nature physics}\ }\textbf {\bibinfo {volume} {12}},\ \bibinfo
  {pages} {826} (\bibinfo {year} {2016})}\BibitemShut {NoStop}%
\bibitem [{\citenamefont {Campbell}\ \emph {et~al.}(2016)\citenamefont
  {Campbell}, \citenamefont {Price}, \citenamefont {Putra}, \citenamefont
  {Vald{\'e}s-Curiel}, \citenamefont {Trypogeorgos},\ and\ \citenamefont
  {Spielman}}]{campbell2016magnetic}%
  \BibitemOpen
  \bibfield  {author} {\bibinfo {author} {\bibfnamefont {D.}~\bibnamefont
  {Campbell}}, \bibinfo {author} {\bibfnamefont {R.}~\bibnamefont {Price}},
  \bibinfo {author} {\bibfnamefont {A.}~\bibnamefont {Putra}}, \bibinfo
  {author} {\bibfnamefont {A.}~\bibnamefont {Vald{\'e}s-Curiel}}, \bibinfo
  {author} {\bibfnamefont {D.}~\bibnamefont {Trypogeorgos}},\ and\ \bibinfo
  {author} {\bibfnamefont {I.}~\bibnamefont {Spielman}},\ }\href
  {https://doi.org/10.1038/ncomms10897} {\bibfield  {journal} {\bibinfo
  {journal} {Nature Communications}\ }\textbf {\bibinfo {volume} {7}},\
  \bibinfo {pages} {10897} (\bibinfo {year} {2016})}\BibitemShut {NoStop}%
\bibitem [{\citenamefont {Qiu}\ \emph {et~al.}(2020)\citenamefont {Qiu},
  \citenamefont {Liang}, \citenamefont {Yang}, \citenamefont {Yang},
  \citenamefont {Tian}, \citenamefont {Xu},\ and\ \citenamefont
  {Duan}}]{qiu2020observation}%
  \BibitemOpen
  \bibfield  {author} {\bibinfo {author} {\bibfnamefont {L.-Y.}\ \bibnamefont
  {Qiu}}, \bibinfo {author} {\bibfnamefont {H.-Y.}\ \bibnamefont {Liang}},
  \bibinfo {author} {\bibfnamefont {Y.-B.}\ \bibnamefont {Yang}}, \bibinfo
  {author} {\bibfnamefont {H.-X.}\ \bibnamefont {Yang}}, \bibinfo {author}
  {\bibfnamefont {T.}~\bibnamefont {Tian}}, \bibinfo {author} {\bibfnamefont
  {Y.}~\bibnamefont {Xu}},\ and\ \bibinfo {author} {\bibfnamefont {L.-M.}\
  \bibnamefont {Duan}},\ }\href {https://doi.org/10.1126/sciadv.aba7292}
  {\bibfield  {journal} {\bibinfo  {journal} {Science Advances}\ }\textbf
  {\bibinfo {volume} {6}},\ \bibinfo {pages} {eaba7292} (\bibinfo {year}
  {2020})}\BibitemShut {NoStop}%
\bibitem [{\citenamefont {Owerre}\ and\ \citenamefont
  {Paranjape}(2015)}]{OWERRE20151}%
  \BibitemOpen
  \bibfield  {author} {\bibinfo {author} {\bibfnamefont {S.}~\bibnamefont
  {Owerre}}\ and\ \bibinfo {author} {\bibfnamefont {M.}~\bibnamefont
  {Paranjape}},\ }\href {https://doi.org/10.1016/j.physrep.2014.09.001}
  {\bibfield  {journal} {\bibinfo  {journal} {Physics Reports}\ }\textbf
  {\bibinfo {volume} {546}},\ \bibinfo {pages} {1} (\bibinfo {year}
  {2015})}\BibitemShut {NoStop}%
\bibitem [{\citenamefont {Eckardt}(2017)}]{eckardt2017colloquium}%
  \BibitemOpen
  \bibfield  {author} {\bibinfo {author} {\bibfnamefont {A.}~\bibnamefont
  {Eckardt}},\ }\href {https://doi.org/10.1103/RevModPhys.89.011004} {\bibfield
   {journal} {\bibinfo  {journal} {Reviews of Modern Physics}\ }\textbf
  {\bibinfo {volume} {89}},\ \bibinfo {pages} {011004} (\bibinfo {year}
  {2017})}\BibitemShut {NoStop}%
\bibitem [{\citenamefont {Osterloh}\ \emph {et~al.}(2002)\citenamefont
  {Osterloh}, \citenamefont {Amico}, \citenamefont {Falci},\ and\ \citenamefont
  {Fazio}}]{osterloh2002scaling}%
  \BibitemOpen
  \bibfield  {author} {\bibinfo {author} {\bibfnamefont {A.}~\bibnamefont
  {Osterloh}}, \bibinfo {author} {\bibfnamefont {L.}~\bibnamefont {Amico}},
  \bibinfo {author} {\bibfnamefont {G.}~\bibnamefont {Falci}},\ and\ \bibinfo
  {author} {\bibfnamefont {R.}~\bibnamefont {Fazio}},\ }\href
  {https://doi.org/10.1038/416608a} {\bibfield  {journal} {\bibinfo  {journal}
  {Nature}\ }\textbf {\bibinfo {volume} {416}},\ \bibinfo {pages} {608}
  (\bibinfo {year} {2002})}\BibitemShut {NoStop}%
\bibitem [{\citenamefont {Eckardt}\ \emph {et~al.}(2005)\citenamefont
  {Eckardt}, \citenamefont {Weiss},\ and\ \citenamefont
  {Holthaus}}]{eckardt2005superfluid}%
  \BibitemOpen
  \bibfield  {author} {\bibinfo {author} {\bibfnamefont {A.}~\bibnamefont
  {Eckardt}}, \bibinfo {author} {\bibfnamefont {C.}~\bibnamefont {Weiss}},\
  and\ \bibinfo {author} {\bibfnamefont {M.}~\bibnamefont {Holthaus}},\ }\href
  {https://doi.org/10.1103/PhysRevLett.95.260404} {\bibfield  {journal}
  {\bibinfo  {journal} {Physical Review Letters}\ }\textbf {\bibinfo {volume}
  {95}},\ \bibinfo {pages} {260404} (\bibinfo {year} {2005})}\BibitemShut
  {NoStop}%
\bibitem [{\citenamefont {Zenesini}\ \emph {et~al.}(2009)\citenamefont
  {Zenesini}, \citenamefont {Lignier}, \citenamefont {Ciampini}, \citenamefont
  {Morsch},\ and\ \citenamefont {Arimondo}}]{zenesini2009coherent}%
  \BibitemOpen
  \bibfield  {author} {\bibinfo {author} {\bibfnamefont {A.}~\bibnamefont
  {Zenesini}}, \bibinfo {author} {\bibfnamefont {H.}~\bibnamefont {Lignier}},
  \bibinfo {author} {\bibfnamefont {D.}~\bibnamefont {Ciampini}}, \bibinfo
  {author} {\bibfnamefont {O.}~\bibnamefont {Morsch}},\ and\ \bibinfo {author}
  {\bibfnamefont {E.}~\bibnamefont {Arimondo}},\ }\href
  {https://doi.org/10.1103/PhysRevLett.102.100403} {\bibfield  {journal}
  {\bibinfo  {journal} {Physical Review Letters}\ }\textbf {\bibinfo {volume}
  {102}},\ \bibinfo {pages} {100403} (\bibinfo {year} {2009})}\BibitemShut
  {NoStop}%
\bibitem [{\citenamefont {Michon}\ \emph {et~al.}(2018)\citenamefont {Michon},
  \citenamefont {Cabrera-Guti{\'e}rrez}, \citenamefont {Fortun}, \citenamefont
  {Berger}, \citenamefont {Arnal}, \citenamefont {Brunaud}, \citenamefont
  {Billy}, \citenamefont {Petitjean}, \citenamefont {Schlagheck},\ and\
  \citenamefont {Gu{\'e}ry-Odelin}}]{michon2018phase}%
  \BibitemOpen
  \bibfield  {author} {\bibinfo {author} {\bibfnamefont {E.}~\bibnamefont
  {Michon}}, \bibinfo {author} {\bibfnamefont {C.}~\bibnamefont
  {Cabrera-Guti{\'e}rrez}}, \bibinfo {author} {\bibfnamefont {A.}~\bibnamefont
  {Fortun}}, \bibinfo {author} {\bibfnamefont {M.}~\bibnamefont {Berger}},
  \bibinfo {author} {\bibfnamefont {M.}~\bibnamefont {Arnal}}, \bibinfo
  {author} {\bibfnamefont {V.}~\bibnamefont {Brunaud}}, \bibinfo {author}
  {\bibfnamefont {J.}~\bibnamefont {Billy}}, \bibinfo {author} {\bibfnamefont
  {C.}~\bibnamefont {Petitjean}}, \bibinfo {author} {\bibfnamefont
  {P.}~\bibnamefont {Schlagheck}},\ and\ \bibinfo {author} {\bibfnamefont
  {D.}~\bibnamefont {Gu{\'e}ry-Odelin}},\ }\href
  {https://doi.org/10.1088/1367-2630/aabc3f} {\bibfield  {journal} {\bibinfo
  {journal} {New Journal of Physics}\ }\textbf {\bibinfo {volume} {20}},\
  \bibinfo {pages} {053035} (\bibinfo {year} {2018})}\BibitemShut {NoStop}%
\bibitem [{\citenamefont {Cooper}\ \emph {et~al.}(2019)\citenamefont {Cooper},
  \citenamefont {Dalibard},\ and\ \citenamefont
  {Spielman}}]{cooper2019topological}%
  \BibitemOpen
  \bibfield  {author} {\bibinfo {author} {\bibfnamefont {N.~R.}\ \bibnamefont
  {Cooper}}, \bibinfo {author} {\bibfnamefont {J.}~\bibnamefont {Dalibard}},\
  and\ \bibinfo {author} {\bibfnamefont {I.~B.}\ \bibnamefont {Spielman}},\
  }\href {https://doi.org/10.1103/RevModPhys.91.015005} {\bibfield  {journal}
  {\bibinfo  {journal} {Reviews of Modern Physics}\ }\textbf {\bibinfo {volume}
  {91}},\ \bibinfo {pages} {015005} (\bibinfo {year} {2019})}\BibitemShut
  {NoStop}%
\bibitem [{\citenamefont {Zheng}\ \emph {et~al.}(2014)\citenamefont {Zheng},
  \citenamefont {Liu}, \citenamefont {Miao}, \citenamefont {Chin},\ and\
  \citenamefont {Zhai}}]{zheng2014strong}%
  \BibitemOpen
  \bibfield  {author} {\bibinfo {author} {\bibfnamefont {W.}~\bibnamefont
  {Zheng}}, \bibinfo {author} {\bibfnamefont {B.}~\bibnamefont {Liu}}, \bibinfo
  {author} {\bibfnamefont {J.}~\bibnamefont {Miao}}, \bibinfo {author}
  {\bibfnamefont {C.}~\bibnamefont {Chin}},\ and\ \bibinfo {author}
  {\bibfnamefont {H.}~\bibnamefont {Zhai}},\ }\href
  {https://doi.org/10.1103/PhysRevLett.113.155303} {\bibfield  {journal}
  {\bibinfo  {journal} {Physical Review Letters}\ }\textbf {\bibinfo {volume}
  {113}},\ \bibinfo {pages} {155303} (\bibinfo {year} {2014})}\BibitemShut
  {NoStop}%
\bibitem [{\citenamefont {Parker}\ \emph {et~al.}(2013)\citenamefont {Parker},
  \citenamefont {Ha},\ and\ \citenamefont {Chin}}]{parker2013direct}%
  \BibitemOpen
  \bibfield  {author} {\bibinfo {author} {\bibfnamefont {C.~V.}\ \bibnamefont
  {Parker}}, \bibinfo {author} {\bibfnamefont {L.-C.}\ \bibnamefont {Ha}},\
  and\ \bibinfo {author} {\bibfnamefont {C.}~\bibnamefont {Chin}},\ }\href
  {https://doi.org/10.1038/nphys2789} {\bibfield  {journal} {\bibinfo
  {journal} {Nature Physics}\ }\textbf {\bibinfo {volume} {9}},\ \bibinfo
  {pages} {769} (\bibinfo {year} {2013})}\BibitemShut {NoStop}%
\bibitem [{\citenamefont {Lim}\ \emph {et~al.}(2008)\citenamefont {Lim},
  \citenamefont {Smith},\ and\ \citenamefont {Hemmerich}}]{lim2008staggered}%
  \BibitemOpen
  \bibfield  {author} {\bibinfo {author} {\bibfnamefont {L.-K.}\ \bibnamefont
  {Lim}}, \bibinfo {author} {\bibfnamefont {C.~M.}\ \bibnamefont {Smith}},\
  and\ \bibinfo {author} {\bibfnamefont {A.}~\bibnamefont {Hemmerich}},\ }\href
  {https://doi.org/10.1103/PhysRevLett.100.130402} {\bibfield  {journal}
  {\bibinfo  {journal} {Physical Review Letters}\ }\textbf {\bibinfo {volume}
  {100}},\ \bibinfo {pages} {130402} (\bibinfo {year} {2008})}\BibitemShut
  {NoStop}%
\bibitem [{\citenamefont {Str{\"a}ter}\ and\ \citenamefont
  {Eckardt}(2015)}]{strater2015orbital}%
  \BibitemOpen
  \bibfield  {author} {\bibinfo {author} {\bibfnamefont {C.}~\bibnamefont
  {Str{\"a}ter}}\ and\ \bibinfo {author} {\bibfnamefont {A.}~\bibnamefont
  {Eckardt}},\ }\href {https://doi.org/10.1103/PhysRevA.91.053602} {\bibfield
  {journal} {\bibinfo  {journal} {Physical Review A}\ }\textbf {\bibinfo
  {volume} {91}},\ \bibinfo {pages} {053602} (\bibinfo {year}
  {2015})}\BibitemShut {NoStop}%
\bibitem [{Sup()}]{SuppMat}%
  \BibitemOpen
  \href@noop {} {\bibinfo {title} {See supplementary materials.}}\BibitemShut
  {Stop}%
\bibitem [{\citenamefont {Weinberg}\ \emph {et~al.}(2015)\citenamefont
  {Weinberg}, \citenamefont {{\"O}lschl{\"a}ger}, \citenamefont {Str{\"a}ter},
  \citenamefont {Prelle}, \citenamefont {Eckardt}, \citenamefont {Sengstock},\
  and\ \citenamefont {Simonet}}]{weinberg2015multiphoton}%
  \BibitemOpen
  \bibfield  {author} {\bibinfo {author} {\bibfnamefont {M.}~\bibnamefont
  {Weinberg}}, \bibinfo {author} {\bibfnamefont {C.}~\bibnamefont
  {{\"O}lschl{\"a}ger}}, \bibinfo {author} {\bibfnamefont {C.}~\bibnamefont
  {Str{\"a}ter}}, \bibinfo {author} {\bibfnamefont {S.}~\bibnamefont {Prelle}},
  \bibinfo {author} {\bibfnamefont {A.}~\bibnamefont {Eckardt}}, \bibinfo
  {author} {\bibfnamefont {K.}~\bibnamefont {Sengstock}},\ and\ \bibinfo
  {author} {\bibfnamefont {J.}~\bibnamefont {Simonet}},\ }\href
  {https://doi.org/10.1103/PhysRevA.92.043621} {\bibfield  {journal} {\bibinfo
  {journal} {Physical Review A}\ }\textbf {\bibinfo {volume} {92}},\ \bibinfo
  {pages} {043621} (\bibinfo {year} {2015})}\BibitemShut {NoStop}%
\bibitem [{\citenamefont {Arimondo}\ \emph {et~al.}(2012)\citenamefont
  {Arimondo}, \citenamefont {Ciampini}, \citenamefont {Eckardt}, \citenamefont
  {Holthaus},\ and\ \citenamefont {Morsch}}]{Arimondo2012}%
  \BibitemOpen
  \bibfield  {author} {\bibinfo {author} {\bibfnamefont {E.}~\bibnamefont
  {Arimondo}}, \bibinfo {author} {\bibfnamefont {D.}~\bibnamefont {Ciampini}},
  \bibinfo {author} {\bibfnamefont {A.}~\bibnamefont {Eckardt}}, \bibinfo
  {author} {\bibfnamefont {M.}~\bibnamefont {Holthaus}},\ and\ \bibinfo
  {author} {\bibfnamefont {O.}~\bibnamefont {Morsch}},\ }\href
  {https://doi.org/10.1016/B978-0-12-396482-3.00010-7} {\bibfield  {journal}
  {\bibinfo  {journal} {Advances in Atomic, Molecular, and Optical Physics}\
  }\textbf {\bibinfo {volume} {61}},\ \bibinfo {pages} {515} (\bibinfo {year}
  {2012})}\BibitemShut {NoStop}%
\bibitem [{\citenamefont {Schollw\"{o}ck}(2011)}]{Schollwoeck2011}%
  \BibitemOpen
  \bibfield  {author} {\bibinfo {author} {\bibfnamefont {U.}~\bibnamefont
  {Schollw\"{o}ck}},\ }\href {https://doi.org/10.1016/j.aop.2010.09.012}
  {\bibfield  {journal} {\bibinfo  {journal} {Annals of Physics}\ }\textbf
  {\bibinfo {volume} {326}},\ \bibinfo {pages} {96} (\bibinfo {year}
  {2011})}\BibitemShut {NoStop}%
\bibitem [{\citenamefont {Goldman}\ and\ \citenamefont
  {Dalibard}(2014)}]{goldman2014periodically}%
  \BibitemOpen
  \bibfield  {author} {\bibinfo {author} {\bibfnamefont {N.}~\bibnamefont
  {Goldman}}\ and\ \bibinfo {author} {\bibfnamefont {J.}~\bibnamefont
  {Dalibard}},\ }\href {https://doi.org/10.1103/PhysRevX.4.031027} {\bibfield
  {journal} {\bibinfo  {journal} {Physical Review X}\ }\textbf {\bibinfo
  {volume} {4}},\ \bibinfo {pages} {031027} (\bibinfo {year}
  {2014})}\BibitemShut {NoStop}%
\bibitem [{\citenamefont {Paeckel}\ \emph {et~al.}(2019)\citenamefont
  {Paeckel}, \citenamefont {K{\"o}hler}, \citenamefont {Swoboda}, \citenamefont
  {Manmana}, \citenamefont {Schollw{\"o}ck},\ and\ \citenamefont
  {Hubig}}]{Spaeckel2019}%
  \BibitemOpen
  \bibfield  {author} {\bibinfo {author} {\bibfnamefont {S.}~\bibnamefont
  {Paeckel}}, \bibinfo {author} {\bibfnamefont {T.}~\bibnamefont {K{\"o}hler}},
  \bibinfo {author} {\bibfnamefont {A.}~\bibnamefont {Swoboda}}, \bibinfo
  {author} {\bibfnamefont {S.~R.}\ \bibnamefont {Manmana}}, \bibinfo {author}
  {\bibfnamefont {U.}~\bibnamefont {Schollw{\"o}ck}},\ and\ \bibinfo {author}
  {\bibfnamefont {C.}~\bibnamefont {Hubig}},\ }\href
  {https://doi.org/10.1016/j.aop.2019.167998} {\bibfield  {journal} {\bibinfo
  {journal} {Annals of Physics}\ }\textbf {\bibinfo {volume} {411}},\ \bibinfo
  {pages} {167998} (\bibinfo {year} {2019})}\BibitemShut {NoStop}%
\bibitem [{\citenamefont {Clark}\ \emph {et~al.}(2016)\citenamefont {Clark},
  \citenamefont {Feng},\ and\ \citenamefont {Chin}}]{clark2016universal}%
  \BibitemOpen
  \bibfield  {author} {\bibinfo {author} {\bibfnamefont {L.~W.}\ \bibnamefont
  {Clark}}, \bibinfo {author} {\bibfnamefont {L.}~\bibnamefont {Feng}},\ and\
  \bibinfo {author} {\bibfnamefont {C.}~\bibnamefont {Chin}},\ }\href
  {https://doi.org/10.1126/science.aaf9657} {\bibfield  {journal} {\bibinfo
  {journal} {Science}\ }\textbf {\bibinfo {volume} {354}},\ \bibinfo {pages}
  {606} (\bibinfo {year} {2016})}\BibitemShut {NoStop}%
\bibitem [{\citenamefont {Shimizu}\ \emph {et~al.}(2018)\citenamefont
  {Shimizu}, \citenamefont {Hirano}, \citenamefont {Park}, \citenamefont
  {Kuno},\ and\ \citenamefont {Ichinose}}]{shimizu2018dynamics}%
  \BibitemOpen
  \bibfield  {author} {\bibinfo {author} {\bibfnamefont {K.}~\bibnamefont
  {Shimizu}}, \bibinfo {author} {\bibfnamefont {T.}~\bibnamefont {Hirano}},
  \bibinfo {author} {\bibfnamefont {J.}~\bibnamefont {Park}}, \bibinfo {author}
  {\bibfnamefont {Y.}~\bibnamefont {Kuno}},\ and\ \bibinfo {author}
  {\bibfnamefont {I.}~\bibnamefont {Ichinose}},\ }\href
  {https://doi.org/10.1088/1367-2630/aad5f9} {\bibfield  {journal} {\bibinfo
  {journal} {New Journal of Physics}\ }\textbf {\bibinfo {volume} {20}},\
  \bibinfo {pages} {083006} (\bibinfo {year} {2018})}\BibitemShut {NoStop}%
\bibitem [{\citenamefont {Pelissetto}\ \emph {et~al.}(2018)\citenamefont
  {Pelissetto}, \citenamefont {Rossini},\ and\ \citenamefont
  {Vicari}}]{pelissetto2018out}%
  \BibitemOpen
  \bibfield  {author} {\bibinfo {author} {\bibfnamefont {A.}~\bibnamefont
  {Pelissetto}}, \bibinfo {author} {\bibfnamefont {D.}~\bibnamefont
  {Rossini}},\ and\ \bibinfo {author} {\bibfnamefont {E.}~\bibnamefont
  {Vicari}},\ }\href {https://doi.org/10.1103/PhysRevB.97.094414} {\bibfield
  {journal} {\bibinfo  {journal} {Physical Review B}\ }\textbf {\bibinfo
  {volume} {97}},\ \bibinfo {pages} {094414} (\bibinfo {year}
  {2018})}\BibitemShut {NoStop}%
\bibitem [{\citenamefont {Coulamy}\ \emph {et~al.}(2017)\citenamefont
  {Coulamy}, \citenamefont {Saguia},\ and\ \citenamefont
  {Sarandy}}]{coulamy2017dynamics}%
  \BibitemOpen
  \bibfield  {author} {\bibinfo {author} {\bibfnamefont {I.~B.}\ \bibnamefont
  {Coulamy}}, \bibinfo {author} {\bibfnamefont {A.}~\bibnamefont {Saguia}},\
  and\ \bibinfo {author} {\bibfnamefont {M.~S.}\ \bibnamefont {Sarandy}},\
  }\href {https://doi.org/10.1103/PhysRevE.95.022127} {\bibfield  {journal}
  {\bibinfo  {journal} {Physical Review E}\ }\textbf {\bibinfo {volume} {95}},\
  \bibinfo {pages} {022127} (\bibinfo {year} {2017})}\BibitemShut {NoStop}%
\bibitem [{\citenamefont {Amaricci}\ \emph {et~al.}(2015)\citenamefont
  {Amaricci}, \citenamefont {Budich}, \citenamefont {Capone}, \citenamefont
  {Trauzettel},\ and\ \citenamefont {Sangiovanni}}]{amaricci2015first}%
  \BibitemOpen
  \bibfield  {author} {\bibinfo {author} {\bibfnamefont {A.}~\bibnamefont
  {Amaricci}}, \bibinfo {author} {\bibfnamefont {J.}~\bibnamefont {Budich}},
  \bibinfo {author} {\bibfnamefont {M.}~\bibnamefont {Capone}}, \bibinfo
  {author} {\bibfnamefont {B.}~\bibnamefont {Trauzettel}},\ and\ \bibinfo
  {author} {\bibfnamefont {G.}~\bibnamefont {Sangiovanni}},\ }\href
  {https://doi.org/10.1103/PhysRevLett.114.185701} {\bibfield  {journal}
  {\bibinfo  {journal} {Physical Review Letters}\ }\textbf {\bibinfo {volume}
  {114}},\ \bibinfo {pages} {185701} (\bibinfo {year} {2015})}\BibitemShut
  {NoStop}%
\bibitem [{\citenamefont {Zhao}\ \emph {et~al.}(2018)\citenamefont {Zhao},
  \citenamefont {Zhang},\ and\ \citenamefont {Jain}}]{zhao2018crystallization}%
  \BibitemOpen
  \bibfield  {author} {\bibinfo {author} {\bibfnamefont {J.}~\bibnamefont
  {Zhao}}, \bibinfo {author} {\bibfnamefont {Y.}~\bibnamefont {Zhang}},\ and\
  \bibinfo {author} {\bibfnamefont {J.}~\bibnamefont {Jain}},\ }\href
  {https://doi.org/10.1103/PhysRevLett.121.116802} {\bibfield  {journal}
  {\bibinfo  {journal} {Physical Review Letters}\ }\textbf {\bibinfo {volume}
  {121}},\ \bibinfo {pages} {116802} (\bibinfo {year} {2018})}\BibitemShut
  {NoStop}%
\bibitem [{\citenamefont {Wu}\ \emph {et~al.}(2021)\citenamefont {Wu},
  \citenamefont {Yang}, \citenamefont {Green}, \citenamefont {Sandvik},\ and\
  \citenamefont {Chamon}}]{wu2021z2}%
  \BibitemOpen
  \bibfield  {author} {\bibinfo {author} {\bibfnamefont {K.-H.}\ \bibnamefont
  {Wu}}, \bibinfo {author} {\bibfnamefont {Z.-C.}\ \bibnamefont {Yang}},
  \bibinfo {author} {\bibfnamefont {D.}~\bibnamefont {Green}}, \bibinfo
  {author} {\bibfnamefont {A.~W.}\ \bibnamefont {Sandvik}},\ and\ \bibinfo
  {author} {\bibfnamefont {C.}~\bibnamefont {Chamon}},\ }\href
  {https://arxiv.org/abs/2103.16625} {\bibfield  {journal} {\bibinfo  {journal}
  {arXiv:2103.16625}\ } (\bibinfo {year} {2021})}\BibitemShut {NoStop}%
\end{thebibliography}

\begin{thebibliography}{8}%
\makeatletter
\providecommand \@ifxundefined [1]{%
 \@ifx{#1\undefined}
}%
\providecommand \@ifnum [1]{%
 \ifnum #1\expandafter \@firstoftwo
 \else \expandafter \@secondoftwo
 \fi
}%
\providecommand \@ifx [1]{%
 \ifx #1\expandafter \@firstoftwo
 \else \expandafter \@secondoftwo
 \fi
}%
\providecommand \natexlab [1]{#1}%
\providecommand \enquote  [1]{``#1''}%
\providecommand \bibnamefont  [1]{#1}%
\providecommand \bibfnamefont [1]{#1}%
\providecommand \citenamefont [1]{#1}%
\providecommand \href@noop [0]{\@secondoftwo}%
\providecommand \href [0]{\begingroup \@sanitize@url \@href}%
\providecommand \@href[1]{\@@startlink{#1}\@@href}%
\providecommand \@@href[1]{\endgroup#1\@@endlink}%
\providecommand \@sanitize@url [0]{\catcode `\\12\catcode `\$12\catcode
  `\&12\catcode `\#12\catcode `\^12\catcode `\_12\catcode `\%12\relax}%
\providecommand \@@startlink[1]{}%
\providecommand \@@endlink[0]{}%
\providecommand \url  [0]{\begingroup\@sanitize@url \@url }%
\providecommand \@url [1]{\endgroup\@href {#1}{\urlprefix }}%
\providecommand \urlprefix  [0]{URL }%
\providecommand \Eprint [0]{\href }%
\providecommand \doibase [0]{https://doi.org/}%
\providecommand \selectlanguage [0]{\@gobble}%
\providecommand \bibinfo  [0]{\@secondoftwo}%
\providecommand \bibfield  [0]{\@secondoftwo}%
\providecommand \translation [1]{[#1]}%
\providecommand \BibitemOpen [0]{}%
\providecommand \bibitemStop [0]{}%
\providecommand \bibitemNoStop [0]{.\EOS\space}%
\providecommand \EOS [0]{\spacefactor3000\relax}%
\providecommand \BibitemShut  [1]{\csname bibitem#1\endcsname}%
\let\auto@bib@innerbib\@empty
\bibitem [{\citenamefont {F{\"o}lling}\ \emph {et~al.}(2007)\citenamefont
  {F{\"o}lling}, \citenamefont {Trotzky}, \citenamefont {Cheinet},
  \citenamefont {Feld}, \citenamefont {Saers}, \citenamefont {Widera},
  \citenamefont {M{\"u}ller},\ and\ \citenamefont {Bloch}}]{folling2007direct}%
  \BibitemOpen
  \bibfield  {author} {\bibinfo {author} {\bibfnamefont {S.}~\bibnamefont
  {F{\"o}lling}}, \bibinfo {author} {\bibfnamefont {S.}~\bibnamefont
  {Trotzky}}, \bibinfo {author} {\bibfnamefont {P.}~\bibnamefont {Cheinet}},
  \bibinfo {author} {\bibfnamefont {M.}~\bibnamefont {Feld}}, \bibinfo {author}
  {\bibfnamefont {R.}~\bibnamefont {Saers}}, \bibinfo {author} {\bibfnamefont
  {A.}~\bibnamefont {Widera}}, \bibinfo {author} {\bibfnamefont
  {T.}~\bibnamefont {M{\"u}ller}},\ and\ \bibinfo {author} {\bibfnamefont
  {I.}~\bibnamefont {Bloch}},\ }\href {https://doi.org/10.1038/nature06112}
  {\bibfield  {journal} {\bibinfo  {journal} {Nature}\ }\textbf {\bibinfo
  {volume} {448}},\ \bibinfo {pages} {1029} (\bibinfo {year}
  {2007})}\BibitemShut {NoStop}%
\bibitem [{\citenamefont {Arimondo}\ \emph {et~al.}(2012)\citenamefont
  {Arimondo}, \citenamefont {Ciampini}, \citenamefont {Eckardt}, \citenamefont
  {Holthaus},\ and\ \citenamefont {Morsch}}]{sArimondo2012}%
  \BibitemOpen
  \bibfield  {author} {\bibinfo {author} {\bibfnamefont {E.}~\bibnamefont
  {Arimondo}}, \bibinfo {author} {\bibfnamefont {D.}~\bibnamefont {Ciampini}},
  \bibinfo {author} {\bibfnamefont {A.}~\bibnamefont {Eckardt}}, \bibinfo
  {author} {\bibfnamefont {M.}~\bibnamefont {Holthaus}},\ and\ \bibinfo
  {author} {\bibfnamefont {O.}~\bibnamefont {Morsch}},\ }\href
  {https://doi.org/10.1016/b978-0-12-396482-3.00010-7} {\bibfield  {journal}
  {\bibinfo  {journal} {Adv. At. Mol. Opt. Phys.}\ }\textbf {\bibinfo {volume}
  {61}},\ \bibinfo {pages} {515} (\bibinfo {year} {2012})}\BibitemShut
  {NoStop}%
\bibitem [{\citenamefont {Drese}\ and\ \citenamefont
  {Holthaus}(1997)}]{sDrese1997}%
  \BibitemOpen
  \bibfield  {author} {\bibinfo {author} {\bibfnamefont {K.}~\bibnamefont
  {Drese}}\ and\ \bibinfo {author} {\bibfnamefont {M.}~\bibnamefont
  {Holthaus}},\ }\href {https://doi.org/10.1016/s0301-0104(97)00025-6}
  {\bibfield  {journal} {\bibinfo  {journal} {Chemical Physics}\ }\textbf
  {\bibinfo {volume} {217}},\ \bibinfo {pages} {201} (\bibinfo {year}
  {1997})}\BibitemShut {NoStop}%
\bibitem [{\citenamefont {Ejima}\ \emph {et~al.}(2012)\citenamefont {Ejima},
  \citenamefont {Fehske}, \citenamefont {Gebhard}, \citenamefont
  {zu~M{\"u}nster}, \citenamefont {Knap}, \citenamefont {Arrigoni},\ and\
  \citenamefont {von~der Linden}}]{sEjima2012}%
  \BibitemOpen
  \bibfield  {author} {\bibinfo {author} {\bibfnamefont {S.}~\bibnamefont
  {Ejima}}, \bibinfo {author} {\bibfnamefont {H.}~\bibnamefont {Fehske}},
  \bibinfo {author} {\bibfnamefont {F.}~\bibnamefont {Gebhard}}, \bibinfo
  {author} {\bibfnamefont {K.}~\bibnamefont {zu~M{\"u}nster}}, \bibinfo
  {author} {\bibfnamefont {M.}~\bibnamefont {Knap}}, \bibinfo {author}
  {\bibfnamefont {E.}~\bibnamefont {Arrigoni}},\ and\ \bibinfo {author}
  {\bibfnamefont {W.}~\bibnamefont {von~der Linden}},\ }\href
  {https://doi.org/10.1103/physreva.85.053644} {\bibfield  {journal} {\bibinfo
  {journal} {Physical Review A}\ }\textbf {\bibinfo {volume} {85}},\ \bibinfo
  {pages} {053644} (\bibinfo {year} {2012})}\BibitemShut {NoStop}%
\bibitem [{\citenamefont {Paeckel}\ \emph {et~al.}(2019)\citenamefont
  {Paeckel}, \citenamefont {K{\"o}hler}, \citenamefont {Swoboda}, \citenamefont
  {Manmana}, \citenamefont {Schollw{\"o}ck},\ and\ \citenamefont
  {Hubig}}]{spaeckel2019time}%
  \BibitemOpen
  \bibfield  {author} {\bibinfo {author} {\bibfnamefont {S.}~\bibnamefont
  {Paeckel}}, \bibinfo {author} {\bibfnamefont {T.}~\bibnamefont {K{\"o}hler}},
  \bibinfo {author} {\bibfnamefont {A.}~\bibnamefont {Swoboda}}, \bibinfo
  {author} {\bibfnamefont {S.~R.}\ \bibnamefont {Manmana}}, \bibinfo {author}
  {\bibfnamefont {U.}~\bibnamefont {Schollw{\"o}ck}},\ and\ \bibinfo {author}
  {\bibfnamefont {C.}~\bibnamefont {Hubig}},\ }\href
  {https://doi.org/10.1016/j.aop.2019.167998} {\bibfield  {journal} {\bibinfo
  {journal} {Annals of Physics}\ }\textbf {\bibinfo {volume} {411}},\ \bibinfo
  {pages} {167998} (\bibinfo {year} {2019})}\BibitemShut {NoStop}%
\bibitem [{\citenamefont {Goldman}\ and\ \citenamefont
  {Dalibard}(2014)}]{sgoldman2014periodically}%
  \BibitemOpen
  \bibfield  {author} {\bibinfo {author} {\bibfnamefont {N.}~\bibnamefont
  {Goldman}}\ and\ \bibinfo {author} {\bibfnamefont {J.}~\bibnamefont
  {Dalibard}},\ }\href {https://doi.org/10.1103/PhysRevX.4.031027} {\bibfield
  {journal} {\bibinfo  {journal} {Physical Review X}\ }\textbf {\bibinfo
  {volume} {4}},\ \bibinfo {pages} {031027} (\bibinfo {year}
  {2014})}\BibitemShut {NoStop}%
\bibitem [{\citenamefont {Rachel}\ \emph {et~al.}(2012)\citenamefont {Rachel},
  \citenamefont {Laflorencie}, \citenamefont {Song},\ and\ \citenamefont
  {Hur}}]{sRachel2012}%
  \BibitemOpen
  \bibfield  {author} {\bibinfo {author} {\bibfnamefont {S.}~\bibnamefont
  {Rachel}}, \bibinfo {author} {\bibfnamefont {N.}~\bibnamefont {Laflorencie}},
  \bibinfo {author} {\bibfnamefont {H.~F.}\ \bibnamefont {Song}},\ and\
  \bibinfo {author} {\bibfnamefont {K.~L.}\ \bibnamefont {Hur}},\ }\href
  {https://doi.org/10.1103/physrevlett.108.116401} {\bibfield  {journal}
  {\bibinfo  {journal} {Physical Review Letters}\ }\textbf {\bibinfo {volume}
  {108}},\ \bibinfo {pages} {116401} (\bibinfo {year} {2012})}\BibitemShut
  {NoStop}%
\bibitem [{\citenamefont {Calabrese}\ and\ \citenamefont
  {Cardy}(2004)}]{sCalabrese2004}%
  \BibitemOpen
  \bibfield  {author} {\bibinfo {author} {\bibfnamefont {P.}~\bibnamefont
  {Calabrese}}\ and\ \bibinfo {author} {\bibfnamefont {J.}~\bibnamefont
  {Cardy}},\ }\href {https://doi.org/10.1088/1742-5468/2004/06/p06002}
  {\bibfield  {journal} {\bibinfo  {journal} {Journal of Statistical Mechanics:
  Theory and Experiment}\ }\textbf {\bibinfo {volume} {2004}},\ \bibinfo
  {pages} {P06002} (\bibinfo {year} {2004})}\BibitemShut {NoStop}%
\end{thebibliography}
\end{document}